\documentclass[
    aps,
    superscriptaddress,
    tightenlines,
    nofootinbib,
    floatfix,
    longbibliography,
    notitlepage,
    twocolumn
]{revtex4-2}

\usepackage{siunitx}
\usepackage[dvipsnames]{xcolor}
\usepackage{listings}
\usepackage{nicefrac}
\usepackage{amssymb}
\usepackage{dsfont}
\usepackage[colorlinks=true, linkcolor=blue]{hyperref}
\usepackage{booktabs}
\usepackage{comment}
\usepackage{graphicx}
\usepackage{physics}
\usepackage{tikz}
    \usetikzlibrary{calc,arrows,matrix,arrows.meta,shapes,decorations,decorations.shapes}
    \tikzstyle{arrow} = [very thick,->,>=stealth]
\usepackage[super]{nth}
\usepackage{placeins}

\definecolor{red}{RGB}{213, 0, 0}
\definecolor{pink}{RGB}{216, 27, 96}
\definecolor{purple}{RGB}{156, 39, 176}
\definecolor{deep purple}{RGB}{81, 45, 168}
\definecolor{indigo}{RGB}{40, 53, 147}
\definecolor{blue}{RGB}{13, 71, 161}
\definecolor{light blue}{RGB}{3, 155, 229}
\definecolor{cyan}{RGB}{0, 188, 212}
\definecolor{teal}{RGB}{0, 137, 123}
\definecolor{green}{RGB}{56, 142, 60}
\definecolor{light green}{RGB}{124, 179, 66}
\definecolor{lime}{RGB}{205, 220, 57}
\definecolor{yellow}{RGB}{255, 215, 20}
\definecolor{amber}{RGB}{255, 143, 0}
\definecolor{orange}{RGB}{239, 108, 0}
\definecolor{deep orange}{RGB}{216, 67, 21}
\definecolor{brown}{RGB}{93, 64, 55}
\definecolor{grey}{RGB}{97, 97, 97}
\definecolor{blue grey}{RGB}{84, 110, 122}
\definecolor{white}{RGB}{255,255,255}
\definecolor{black}{RGB}{0,0,0}

\newcommand{\Nbst}{\ensuremath{\mathrm{N}_\mathrm{bst}}}
\newcommand{\Nexp}{\ensuremath{\mathrm{N}_\mathrm{states}}}
\newcommand{\Nparams}{\ensuremath{\mathrm{N}_\mathrm{params}}}
\newcommand{\Ncfg}{\ensuremath{\mathrm{N}_\mathrm{conf}}}
\newcommand{\Nt}{\ensuremath{\mathrm{N}_\mathrm{t}}}
\newcommand{\Nx}{\ensuremath{\mathrm{N}_\mathrm{x}}}

\newcommand{\AIC}{\ensuremath{\mathrm{AIC}}}
\newcommand{\Csp}{\ensuremath{\mathrm{C}^\mathrm{sp}}}

\newcommand{\Csh}{\ensuremath{\mathrm{C}^\mathrm{sh}}}

\newcommand{\im}{\ensuremath{\mathrm{i}}}

\newcommand{\perylene}{\mathrm{C}_{20}\mathrm{H}_{12}}
\newcommand{\Z}{\ensuremath{\mathcal{Z}}}

\renewcommand{\O}[1]{
    \ifthenelse { \equal {#1} {} }{
        \ensuremath{\mathcal{O}}
    }{
        \ensuremath{\mathcal{O}\left[#1\right]}
    }
}
\renewcommand{\S}[2][]{
    \ifthenelse { \equal {#1} {} }{
        \ensuremath{\mathrm{S}\left[#2\right]}
    }{
        \ensuremath{\mathrm{S}_\mathrm{#1}\left[#2\right]}
    }
}

\newcommand{\CITE}[1]{%
    \ifthenelse { \equal{#1}{} }{%
        \textcolor{fzjred}{\textbf{[Cite]}}%
    }{%
        \textcolor{fzjred}{\textbf{[Cite:~#1]}}%
    }%
}

\newcommand{\jsc}{
    J\"{u}lich Supercomputing Center (JSC), 
    Forschungszentrum J\"{u}lich, 52428 J\"{u}lich, Germany
}

\newcommand{\casa}{
	Center for Advanced Simulation and Analytics (CASA),
	Forschungszentrum J\"{u}lich, 52428 J\"{u}lich, Germany
}
\newcommand{\ias}{
	Institute for Advanced Simulation 4 (IAS-4),
	Forschungszentrum J\"{u}lich, 52428 J\"{u}lich, Germany
}

\newcommand{\bonn}{
    Helmholtz-Institut f\"{u}r Strahlen- und Kernphysik,
    Rheinische Friedrich-Wilhelms-Universit\"{a}t Bonn, 53115 Bonn, Germany
}

\newcommand{\liverpool}{Department of Mathematical Sciences,
	University of Liverpool, Liverpool, L69 7ZL, United Kingdom
}

\date{\today}

\begin{document}

\title{Single Particle Spectrum of Doped $\perylene$-Perylene}

\affiliation{\jsc}
\affiliation{\ias}
\affiliation{\casa}
\affiliation{\bonn}
\affiliation{\liverpool}

\author{Marcel Rodekamp}
    \affiliation{\jsc}
	\affiliation{\casa}
    \affiliation{\bonn}

\author{Evan Berkowitz}
    \affiliation{\jsc}
    \affiliation{\ias}
	\affiliation{\casa}

\author{Christoph Gäntgen}
	\affiliation{\ias}
	\affiliation{\casa}
	\affiliation{\bonn}

\author{Stefan Krieg}
    \affiliation{\jsc}
	\affiliation{\casa}
    \affiliation{\bonn}

\author{Thomas Luu}
    \affiliation{\ias}
    \affiliation{\bonn}

\author{Johann Ostmeyer}
    \affiliation{\bonn}
    \affiliation{\liverpool}

\author{Giovanni Pederiva}
    \affiliation{\jsc}
	\affiliation{\casa}

\begin{abstract}
    We present a Hamiltonian Monte Carlo study of doped perylene $\perylene$ described with the Hubbard model.
    Doped perylene can be used for organic light-emitting diodes (OLEDs) or as acceptor material in organic solar cells.
    Therefore, central to this study is a scan over charge chemical potential.
    A variational basis of operators allows for the extraction of the single-particle spectrum through a mostly automatic fitting procedure.
    Finite chemical potential simulations suffer from a sign problem which we ameliorate through contour deformation.
    The on-site interaction is kept at $\nicefrac{U}{\kappa}=2$.
    Discretization effects are handled through a continuum limit extrapolation. 
    Our first-principles calculation shows significant deviation from non-interacting results especially at large chemical potentials.
\end{abstract}

\maketitle

\section{Introduction}\label{sec:introduction}
The perylene molecule $\perylene$, pictured in Fig.~\ref{fig:perylene}, has attracted great interest in various technological applications, ranging from
organic semiconductors~\cite{dodabalapur1996molecular,guo2017effect}, 
organic light emitting diodes (OLEDs)~\cite{sato1998operation}, 
to organic solar cells~\cite{tang1986twolayer,ni2021ultrafast,cao2022progress}.
As it is a polycyclic aromatic hydrocarbon, it is also of great interest to astronomy; perylene and its derivatives have been found in interstellar gases and nebulae~\cite{halasinski2003electronic,salama2008pahs,li2020spitzer}.

The ionization energy and electron affinity of perylene is well studied experimentally~\cite{shchuka1989TwoPhoton,schiedt1997Photodetachment}.
Kinetic Monte Carlo simulations have also been conducted involving ensembles of perylene molecules, see e.g.\@~\cite{manian2021singlet,davino2022electron}. 
Theoretical studies of the electronic structure of perylene have been performed using various methods, for example density functional theory (DFT)~\cite{halasinski2003electronic,clark2007exciton} and DMRG~\cite{giri2018model}.

In derivatives of perylene the $\pi$ orbitals of the $\mathrm{sp}^2$-hybridized valence orbitals will not be half-filled; additional bonded groups may supply or draw away electrons.
However, to our knowledge, little is theoretically known about the electronic structure of a single doped perylene molecule.
We therefore model perylene's $\pi$ electrons using the Hubbard model and perform ab-intio grand-canonical Monte Carlo simulations to map the single-electron spectrum as a function of the electron chemical potential $\mu$.
We describe this model in Sec.~\ref{ssec:modelling_perylene}.

We describe our computational approach in Sec.~\ref{ssec:simulation_methods}.
In particular, at non-zero $\mu$ our system is not half-filled and our simulations are afflicted by a numerical sign problem. 
We briefly describe the issue and how we leverage recent developments to nevertheless get reliable statistical estimates~\cite{mukherjee2014lefschetz,tanizaki2016lefschetzthimble,ulybyshev2019taming,ulybyshev2020lefschetz}.

We measure the global charge and single-particle (and single-hole) euclidean-time correlation functions from which we extract energy spectra.
In section~\ref{sec:analysis} we explain how this analysis is performed but relegate many details to Appendix~\ref{apx-sec:analysis_details} and further results to Appendix~\ref{apx-sec:moreSpectrum}.
Finally, we summarize our findings in section~\ref{sec:conclusions}.

\section{Formalism}\label{sec:formalism}
\subsection{Modelling Perylene}\label{ssec:modelling_perylene}
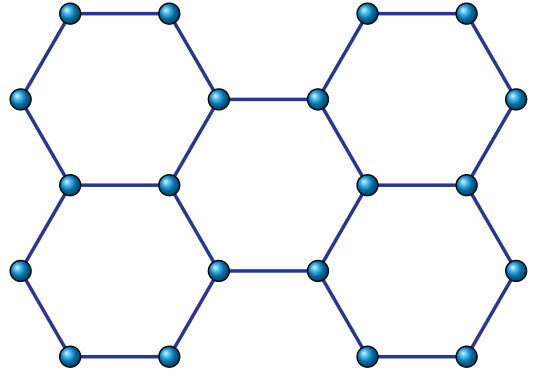
\begin{figure}
    \centering
    \resizebox{0.4\textwidth}{!}{
        \begin{tikzpicture}
            \coordinate (N01) at (0,0);
            \coordinate (N02) at (1,0);
            \coordinate (N03) at (3,0);
            \coordinate (N04) at (4,0);
            \coordinate (N05) at (4.5,-0.8660254037844386);
            \coordinate (N06) at (2.5,-0.8660254037844386);
            \coordinate (N07) at (1.5,-0.8660254037844386);
            \coordinate (N08) at (-0.5,-0.8660254037844386);
            \coordinate (N09) at (0,-1.73205080756888);
            \coordinate (N10) at (1,-1.73205080756888);
            \coordinate (N11) at (3,-1.73205080756888);
            \coordinate (N12) at (4,-1.73205080756888);
            \coordinate (N13) at (4.5,-2.59807621135332);
            \coordinate (N14) at (2.5,-2.59807621135332);
            \coordinate (N15) at (1.5,-2.59807621135332);
            \coordinate (N16) at (-0.5,-2.59807621135332);
            \coordinate (N17) at (0,-3.46410161513775);
            \coordinate (N18) at (1,-3.46410161513775);
            \coordinate (N19) at (3,-3.46410161513775);
            \coordinate (N20) at (4,-3.46410161513775);
        
            \draw[indigo,line width = 1px] (N01) -- (N02);
            \draw[indigo,line width = 1px] (N02) -- (N07);
            \draw[indigo,line width = 1px] (N07) -- (N06);
            \draw[indigo,line width = 1px] (N06) -- (N03);
            \draw[indigo,line width = 1px] (N03) -- (N04);
            \draw[indigo,line width = 1px] (N04) -- (N05);
            \draw[indigo,line width = 1px] (N05) -- (N12);
            \draw[indigo,line width = 1px] (N12) -- (N13);
            \draw[indigo,line width = 1px] (N13) -- (N20);
            \draw[indigo,line width = 1px] (N20) -- (N19);
            \draw[indigo,line width = 1px] (N19) -- (N14);
            \draw[indigo,line width = 1px] (N14) -- (N15);
            \draw[indigo,line width = 1px] (N15) -- (N18);
            \draw[indigo,line width = 1px] (N18) -- (N17);
            \draw[indigo,line width = 1px] (N17) -- (N16);
            \draw[indigo,line width = 1px] (N16) -- (N09);
            \draw[indigo,line width = 1px] (N09) -- (N08);
            \draw[indigo,line width = 1px] (N08) -- (N01);
            \draw[indigo,line width = 1px] (N09) -- (N10);
            \draw[indigo,line width = 1px] (N10) -- (N07);
            \draw[indigo,line width = 1px] (N10) -- (N15);
            \draw[indigo,line width = 1px] (N12) -- (N11);
            \draw[indigo,line width = 1px] (N11) -- (N06);
            \draw[indigo,line width = 1px] (N11) -- (N14);
        
            \draw[ball color = light blue] (N01) circle (3pt);
            \draw[ball color = light blue] (N02) circle (3pt);
            \draw[ball color = light blue] (N03) circle (3pt);
            \draw[ball color = light blue] (N04) circle (3pt);
            \draw[ball color = light blue] (N05) circle (3pt);
            \draw[ball color = light blue] (N06) circle (3pt);
            \draw[ball color = light blue] (N07) circle (3pt);
            \draw[ball color = light blue] (N08) circle (3pt);
            \draw[ball color = light blue] (N09) circle (3pt);
            \draw[ball color = light blue] (N10) circle (3pt);
            \draw[ball color = light blue] (N11) circle (3pt);
            \draw[ball color = light blue] (N12) circle (3pt);
            \draw[ball color = light blue] (N13) circle (3pt);
            \draw[ball color = light blue] (N14) circle (3pt);
            \draw[ball color = light blue] (N15) circle (3pt);
            \draw[ball color = light blue] (N16) circle (3pt);
            \draw[ball color = light blue] (N17) circle (3pt);
            \draw[ball color = light blue] (N18) circle (3pt);
            \draw[ball color = light blue] (N19) circle (3pt);
            \draw[ball color = light blue] (N20) circle (3pt);

        \end{tikzpicture}
    }
    \caption{
    Graphical representation of the perylene molecule. 
    The sites represent carbon-ions, while links indicate allowed hopping.
    External hydrogen atoms are not drawn.
    }\label{fig:perylene}
\end{figure}
Perylene consists of $\mathrm{sp}^2$-hybridized carbon atoms arranged in five hexagons~\cite{donaldson1953crystal,botoshansky2003complete}, giving \Nx=20 ions as shown in figure~\ref{fig:perylene}, and twelve hydrogen atoms bonded to the carbons on the boundary (which are not shown in fig.~\ref{fig:perylene}).
The hybridized nature of the carbon bonds allows the valence $\pi$ electrons to hop along the bonds.
We model the kinematics and interactions of these $\pi$-electrons with the Hubbard model
\begin{equation}
\begin{aligned}
    \mathcal{H}\left[\kappa, U, \mu\right]
    =
         &-\kappa \sum_{\langle x,y\rangle\in X} \left( p_x^\dagger p_y^{\phantom{\dagger}} - h_x^\dagger  h_y^{\phantom{\dagger}} \right)\\ %
         &+ \frac{U}{2} \sum_{x\in X} q_x^2
         - \mu \sum_{x\in X} q_x^{\phantom{2}}.
    \label{eq:hubbard-hamiltonian}
\end{aligned}
\end{equation}
The hopping strength $\kappa$ (which we take to be bond-independent) is the amplitude for a free electron to traverse the bond between nearest neighbors $\langle x,y\rangle$.
We work in the particle/hole basis for computational reasons~\cite{wynen2019avoiding}; the $p_x$ ($h_x$) represents a particle (hole) annihilation operator.
We denote the collection of ions by X.
The strength of interaction depends on the charge per site $q_x = h^\dagger_x h_x^{\phantom{\dagger}} - p^\dagger_x p_x^{\phantom{\dagger}}$ (so that particles represent electrons with negative electric charge), and is controlled by the onsite term $U$; a more realistic two-body interaction $\sum_{xy} q_x V_{xy} q_y$ can be easily incorporated into our simulations.

Typical applications of perylene involve attaching additional chemical structures to a perylene core~\cite{zhao2008theoretical,lai2015synthesis}.
To model the electrons in these chemical derivatives in our simulations, we apply a homogeneous effective chemical potential $\mu$ coupling to the total system charge. 
For simplicity, we will provide all physical quantities in units of the hopping strength, i.e.\ $\nicefrac{U}{\kappa}$, $\nicefrac{\mu}{\kappa}$, $E/\kappa$, etc.
and in what follows, we will express these quantities already rescaled by $\kappa$.
Following~\cite{giri2018model}, we can reintroduce physical units setting $\kappa = \SI{2.4}{\eV}$.

The point symmetry group of perylene is typically identified as $D_{2h}$.
Our Hamiltonian \eqref{eq:hubbard-hamiltonian}, however, treats the ions as a fixed graph with no knowledge of its three-dimensional embedding, and we can split the symmetry into the dihedral group $D_2$ and a $\mathbb{Z}_2$ whose only action is to flip spin components (which amounts to an exchange of particles and holes).
Hamiltonian eigenstates will have definite spin and will transform in the $A$, $B_1$, $B_2$, and $B_3$ representations of $D_2$, which are all one-dimensional.

We can perform a basis transformation of the 20 single-particle position-space operators.
The vector space defined on the 20 sites can be decomposed into invariant subspaces on which the action of the $D_2$ symmetries act irreducibly as $A$, $B_1$, $B_2$, and $B_3$; in a slight but common abuse of language we identify these invariant subspaces as the irreps themselves.
The irreps have multiplicity 6, 4, 6, and 4, respectively.

We can arrange for this transformation to diagonalize the hopping matrix  $K=\kappa \delta_{\langle x, y \rangle}$
These operators are shown in detail in Appendix~\ref{subsec:Diagonalize Correlators}; each operator has definite irrep and tight-binding energy $\epsilon$.
In the non-interacting $U=0$ case these irreducible operators carry definite energy and satisfy $[H, p^\dagger_{\Lambda_i}] = \epsilon^{\phantom{\dagger}}_{\Lambda_i} p^\dagger_{\Lambda_i}$ where the state is labelled by irrep $\Lambda$ and an index $i$.
The same transformation can be made to the holes; the only difference arises from the sign of the hopping term for the holes in the Hamiltonian \eqref{eq:hubbard-hamiltonian}.
Some operators have positive tight-binding energy and others have negative tight-binding energy; in the non-interacting case the global ground state consists of every negative-energy operator applied to the Fock vacuum.
 
\subsection{Simulation Methods}\label{ssec:simulation_methods}
We compute observables $\O{}$ expressed through the thermal trace over all Fock space states,
\begin{equation}
    \expval{\O{}} = \frac{1}{\Z} \Tr{ \O{} e^{-\beta \mathcal{H}} }\label{eq:thermal-trace}\ .
\end{equation}
Here the partition function $\Z = \Tr{e^{-\beta \mathcal{H}}}$ and $\beta = \nicefrac{1}{T}$ is the inverse temperature in natural units, $c = k_B = \hbar = 1$.
We Trotterize $\beta$ into $N_t$ timeslices each separated by the temporal lattice spacing $\delta = \nicefrac{\beta}{\Nt}$.
We introduce a continuous auxiliary field $\Phi$ on every site of the spacetime lattice via a Hubbard-Stratonovich transformation~\cite{brower2012hybrid,smith2014montecarlo,ulybyshev2013montecarlo,luu2016quantum} $\Phi = (\Phi_{tx}) \in \mathbb{R}^\abs{\Lambda}$, with indices on the spacetime lattice $ \Lambda = \left[0,\Nt-1\right] \otimes X$.
Exactly integrating out the fermions transforms our problem from a discrete sum over Fock states into a path integral~\cite{hubbard1959calculation,hubbard1963electronI,hubbard1964electronII,hubbard1964electronIII, ostmeyer2020semimetal, wynen2021leveraging},
\begin{equation}\label{eq:expectationValue}
    \expval{\O{}} = \frac{1}{\Z} \int\mathcal{D}\left[\Phi\right] e^{-S\left[\Phi\right]} \O{\Phi}
\end{equation}
where the action $S$ is
\begin{align}
    \label{eq:action}
    \S{\Phi \, \vert \, \kappa,U,\mu} = \frac{ \Phi^2 }{2 \delta U}
    &-\log\det{ M\left[\phantom{\textrm{-}}\Phi \, \vert \, \phantom{\textrm{-}}\kappa,\phantom{\textrm{-}}\mu\right] } \\
    &-\log\det{ M\left[\textrm{-} \Phi \, \vert \, \textrm{-}\kappa,\textrm{-}\mu\right] }\,, \nonumber
\end{align}
and the Gaussian piece can be replaced by $\nicefrac{1}{2}\; \Phi (\delta V)^{-1} \Phi$ for a more generic interaction, as long as the interaction matrix $V_{xy}$ is positive definite.
The fermion matrices are in the exponential discretization~\cite{wynen2019avoiding} 
\begin{align}\label{eq:fermion-matrix}
    M\left[\Phi\,\vert\, K,\mu\right]_{x't';xt} 
    &= 
        \delta_{x'x}\delta_{t't}
    \\
    &-  \left( e^{\delta(K - \mu)} \right)_{x'x} e^{+ i \Phi_{xt}} \mathcal{B}_{t'}\delta_{t'(t+1)}
\nonumber
\end{align}
where $\mathcal{B}$ encodes the anti-periodic boundary conditions in time.
We perform the path integral stochastically using the Hybrid/Hamilton Monte Carlo (HMC) algorithm~\cite{duane1987hybrid}.

At finite chemical potential the fermionic part of the action $\mathrm S$ can become complex, and removes any ergodicity problem~\cite{wynen2019avoiding}.
However, it also introduces the so-called `sign problem' since $e^{-\mathrm S}$ can oscillate. 
A severe sign problem ultimately results in unreliable statistical estimates of observables with finite statistics. 

Complex actions and integrand oscillations can arise across a wide set of computational models and approaches, ranging across $\phi^4$ theory~\cite{cristoforetti2013monte,Fujii2013Hybrid}, topological (Chern-Simons) models~\cite{kanazawa2015structure}, molecular systems~\cite{zhang2017} and lattice QCD~\cite{splittorff2007phase,cristoforetti2012new}, for example.
In recent years there has been a great push to leverage contour deformation to mitigate the sign problem in all these theories.
In addition to trying to deform the contour integration onto Lefschetz thimbles~\cite{cristoforetti2014efficient,alexandru2016fast,alexandru2016monte,fukuma2019applying,fukuma2020implementation}, machine learning methods~\cite{alexandru2016monte,mori2018application,kashiwa2019application,lawrence2019thimbles,alexandru2022complex} can often but not always~\cite{Lawrence:2023sfc} locate integration contours with much more modest problems.
Related deformations to complex Langevin methods~\cite{berger2021complex,fujisawa2022backpropagating} are also undergoing rapid development. 
Moreover, the signal-to-noise problem present for many observables in Markov Chain Monte Carlo simulations can be improved with a similar approach~\cite{detmold2020path,detmold2021path}.

Leveraging experience gained while developing these methods for the Hubbard model~\cite{wynen2019avoiding,wynen2021leveraging,rodekamp2022complex,rodekamp2023mitigatingNovel,gantgen2024fermionic,gantgen2024reducing,rodekamp2024theory}, we perform a simple and cost-efficient transformation by incorporating a spacetime constant imaginary shift $\phi_c$
\begin{equation}
    \Psi(\Phi) = \Phi + \im \phi_c.
    \label{eq:constant-shift}
\end{equation}
Such a shift represents an integration manifold in the complex plane that is parallel to the real plane.
For this investigation we utilize the next-to-leading order (NLO) plane~\cite{gantgen2024fermionic}, whereby $\phi_c$ is determined by including quantum (thermal) corrections to the saddle-point approximation of $S$.
We briefly motivate this method in Appendix~\ref{apx-sec:complex_contour}.
Even with this shift in the integration contour the action remains complex and we perform HMC changing the real part of $\Phi$ according to the real part of the HMC force, accepting proposed changes according to the real part of the action,  and reweighting with the imaginary part of the action as described in Appendix~\ref{subsec:Reweighting}.

\section{Analysis}\label{sec:analysis}
The goal of this investigation is to assess the single particle spectrum in relation to the system's total charge, as a measure of doping.
These two quantities can be obtained by calculating the euclidean time single particle ($p$) and hole ($h$) correlators
\begin{equation}
\begin{alignedat}{2}
    \label{eq:correlator-Minv}
    \Csp_{x,y}\left(\tau\right) &= \expval{ p^{\phantom{\dagger}}_x(\tau) p^\dagger_y(0) } &&= \expval{M^{-1}_{x,\tau;y,0}\left[\phantom{\textrm{-}}\Phi\vert \phantom{\textrm{-}}\kappa,\phantom{\textrm{-}}\mu\right]}\,,\\
    \Csh_{x,y}\left(\tau\right) &= \expval{ h^{\phantom{\dagger}}_x(\tau) h^\dagger_y(0) } &&= \expval{M^{-1}_{x,\tau;y,0}\left[\textrm{-}\Phi\vert \textrm{-}\kappa,\textrm{-}\mu\right]} \,,
\end{alignedat}
\end{equation}
which we can analyze using the standard spectral decomposition (Appendix~\ref{subsec:effective masses}).

After averaging particles and time-reversed holes we have a $20\times20$ matrix of correlators for each ensemble.
The irreducible representation is a good quantum number, allowing us to block-diagonalize to four small correlators, one for each $A$ ($6 \times 6$), $B_1$ ($4\times 4$), $B_2$ ($6 \times 6$), and $B_3$ ($4\times 4$) using the irreducible single-particle operators.
Interactions can mix the operators within an irrep and we variationally extract the six or four interacting energy levels closest to the fully interacting ground state as explained in Appendix~\ref{subsec:Diagonalize Correlators}.

The chemical potential $\mu$ controls the total charge of the system.
To quantify its effect, we compute the total system charge by
\begin{equation}
\begin{aligned}
	\expval{Q}
	= \sum_{x\in X} \expval{q^{\phantom{\dagger}}_x}
	&= \sum_{x} \left(\expval{ h^\dagger_x h^{\phantom{\dagger}}_x - \expval{ p^\dagger_x p^{\phantom{\dagger}}_x } } \right)
	\\
    &=\sum_{x} \left(\expval{ p^{\phantom{\dagger}}_x p_x^\dagger } - \expval{ h_x^{\phantom{\dagger}} h_x^\dagger } \right)
	\\
    &=\sum_{x} \left(\Csp_{x,x}(0) - \Csh_{x,x}(0)\right)\ ,
    \label{eq:charge}
\end{aligned}
\end{equation}
as a function of $\mu$.

In the non-interacting case we can compute the total charge
\begin{align}
    \expval{Q} \vert_{U=0} &= 2 \sum_{ \Lambda_i } \frac{ 1 }{ e^{-\beta (\epsilon_{\Lambda_i}+\mu) } + 1} - \Nx, \label{eq:QFiniteTemp}\\
    \lim_{\beta\to\infty} \expval{Q} \vert_{U=0} &= 2 \sum_{ \Lambda_i } \Theta\left( \epsilon_{\Lambda_i}+\mu \right) - \Nx. \label{eq:QZeroTemp}
\end{align}
The factor of two comes from the spin degeneracy and the subtraction by $N_x$ ensures that $Q=0$ when $\mu=0$.

At non-zero interaction, $U \neq 0$, observables are computed using the NLO-plane HMC algorithm as discussed in the previous section.
This alleviates the sign problem sufficiently to allow us to extract statistically meaningful quantities.
Further details on the analysis steps can be found in appendix~\ref{apx-sec:analysis_details}.

We perform our studies using an on-site interaction of $U = 2$.
This provides us with an initial qualitative behavior of perylene's charge $Q$ as a function of $\mu$.
In the future we aim to tune this on-site coupling to a more realistic value or use a more realistic two-body interaction.

To access different total charges, we scan over the chemical potential $\mu = 0, 0.1,\,\dots, 1.1$.
This choice is inspired by the non-interacting charges discussed in section~\ref{sub-sec:charge}.
We control the temporal continuum limit using three time discretizations $\Nt = 32,64,96$ and study the temperature dependence with $\beta = 4, 6, 8$.
For each parameter combination we measure a total of $\Ncfg = \num{10000}$ configurations.

\subsection{Statistical Power}\label{sub-sec:statistical_power}
Before discussing the analysis of the physical observables, i.e.~\eqref{eq:correlator-Minv} and~\eqref{eq:charge}, it is important to map out the severity of the sign problem.
A typical measure is the absolute average phase, called the statistical power,
\begin{equation}
    \abs{\expval{\Sigma}} = \abs{\expval{ e^{-\im \Im{ \S{\Phi}}} }}.
    \label{eq:statistical power}
\end{equation}
A value of 1 for the statistical power implies no sign problem, whereas a value of 0 represents the most severe sign problem.  
One can further relate the statistical power to an effective number of configurations $\Ncfg^\mathrm{eff} \propto \abs{\expval{\Sigma}}^2 \Ncfg$~\cite{berger2021complex}; when the statistical power is small each configuration is worth less.
The average phase appears in the denominator when reweighting (Appendix~\ref{subsec:Reweighting}) and, therefore, for small, hard-to-estimate statistical powers, stochastic estimates of observables become unreliable.

\begin{figure}[!ht]
    \includegraphics[width = 1\linewidth]{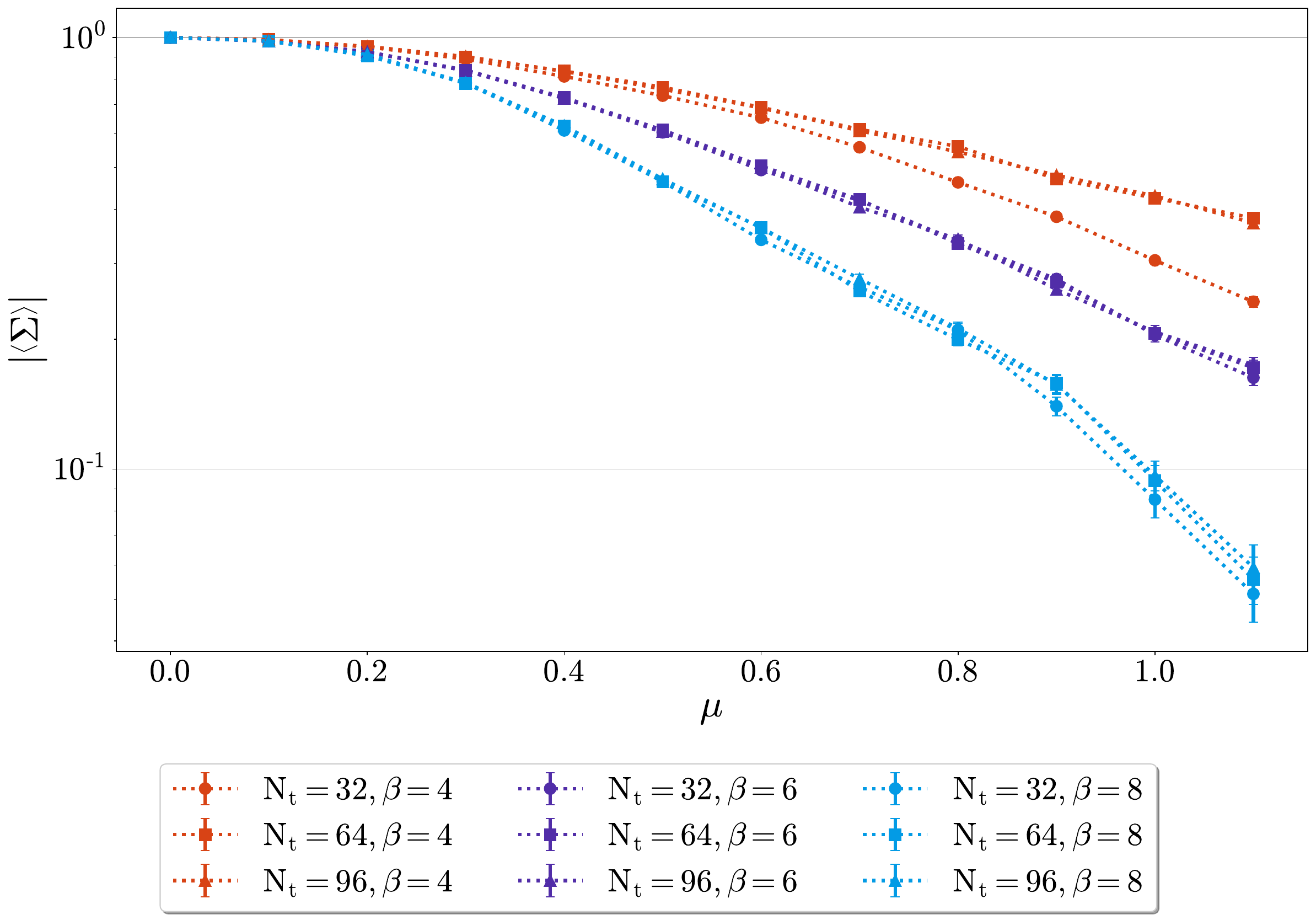}
    \caption{
        Statistical power $\abs{\expval{\Sigma}}$ as a function of the chemical potential $\mu$.
        With the given amount of configurations, beyond $\abs{\expval{\Sigma}} \lessapprox 0.1$ simulations are unreliable.
    }\label{fig:stat-power-per-mu}
\end{figure}

In figure~\ref{fig:stat-power-per-mu} we show the statistical power as a function of $\mu$ plotted for the various $\beta$ and $\Nt$.
With $\Ncfg = \num{10000}$ configurations, simulations with $\abs{\expval{\Sigma}} \lesssim 0.1$ become unreliable.
We emphasize that without the contour deformation \eqref{eq:constant-shift} the statistical power is indistinguishable from 0 for almost all of the $\mu\neq0$ ensembles shown.

We observe that the total system charge (Sec.~\ref{sub-sec:charge}) is less susceptible to statistical noise which allows us to access it over all considered chemical potentials.
In contrast %
the single particle energy spectrum (Sec.~\ref{sub-sec:spectrum}) is more susceptible to the noise resulting in significant uncertainty at $\beta = 8$ with $\mu = 0.9, 1$.
At $\mu = 1.1$ more data is required to reliably estimate the larger energies.
Consequently, we remove this point from the analysis.

\subsection{Total System Charge}\label{sub-sec:charge}
\begin{figure}[h!]
    \includegraphics[width = 1\linewidth]{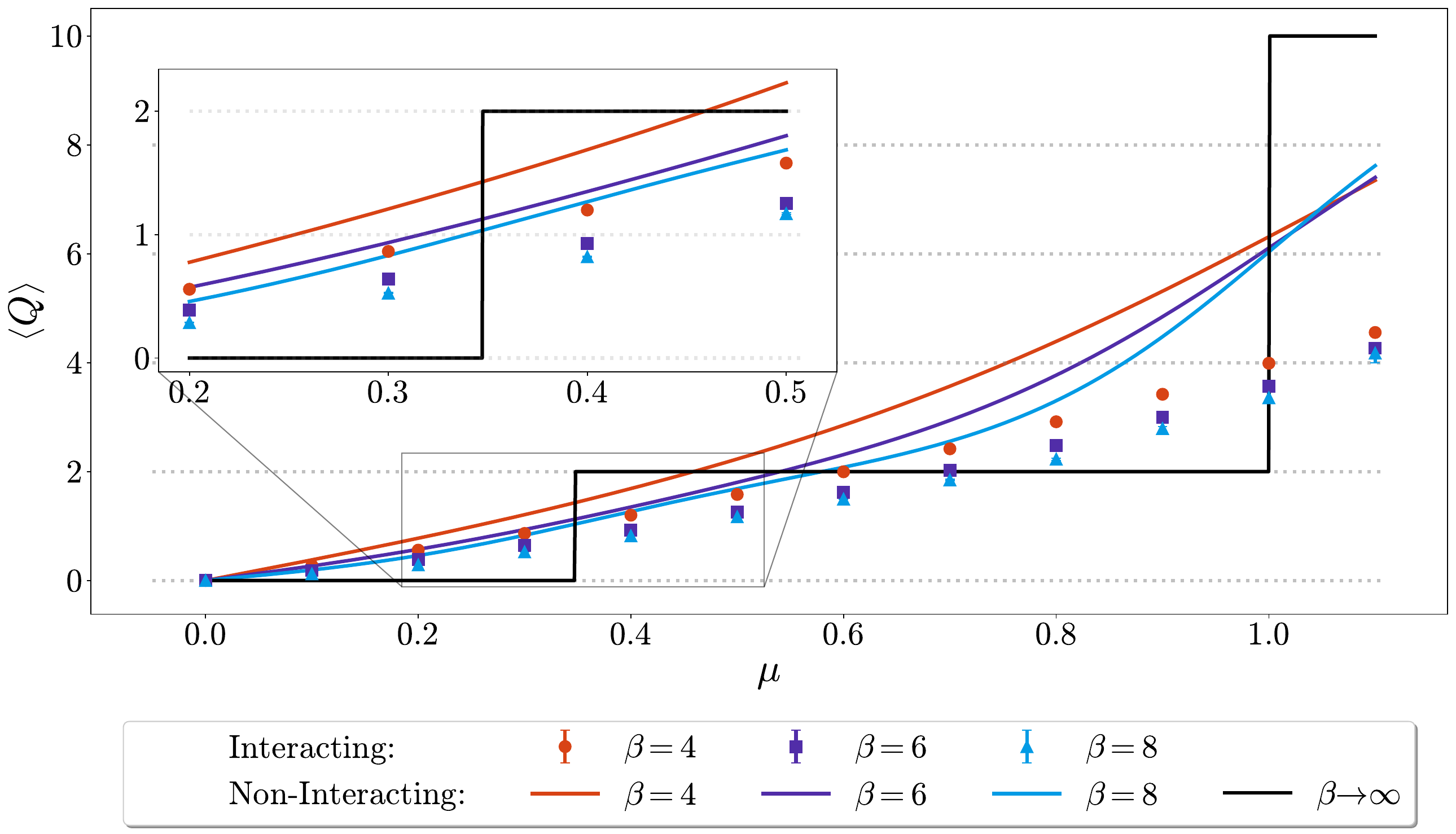}
    \caption{
        Plot of the total system charge as a function of the chemical potential $\mu$.
    }\label{fig:charge-per-mu}
\end{figure}

In figure~\ref{fig:charge-per-mu} we show the charge as a function of $\mu$.
Solid lines are exact non-interacting $U=0$ results, while  the continuum-extrapolated $(\delta\rightarrow 0)$ total system charge measured with $U=2$ is shown as points with uncertainties.
As shown in appendix~\ref{subsec:Limits} our data is close to the continuum limit and we simply fit a constant to the charge at all three $\Nt$ values.

Focusing on the non-interacting result, colored lines are at the simulated $\beta$ while the black line represents the zero-temperature limit.
For the latter we observe a sudden jump of $\Delta Q = 2$ at $\mu\approx 0.347$ which corresponds to the smallest single particle energy $\Delta E_{U = 0}^{B_3^3}(\mu = 0)$; the jump corresponds to two electrons moving out of the Fermi sea.
The difference in charge must be a multiple of two due to the spin-degeneracy preserved in the Hubbard model. %
In the non-interacting case a further jump of $\Delta Q = 8$ appears at $\mu= 1$ corresponding to the next (accidentally-quadruply-degenerate) single particle energy.
Corresponding single particle energies can be found in the first panel of figure~\ref{fig:spectrum}.

Finite temperature washes out the step function \eqref{eq:QFiniteTemp} and we draw colored solid lines for each temperature we simulated.
They cross the $\Delta Q = 2$ threshold necessarily at higher chemical potentials due to finite temperature effects. 
Furthermore, at finite temperature $Q\neq0$ states are partly populated and we can cross $Q=1$ below the free zero-temperature single-particle threshold.

The circles, squares and triangles in figure~\ref{fig:charge-per-mu} display the continuum limit of the charge at finite temperature, $\beta=4,6,8$ respectively. 
The temperatures are too hot to identify a clear charge jump, however, the $\beta = 6,8$ data go through $Q=1$ between $\mu = 0.4 \text{ and } 0.5$, later than the free system.

Comparing the finite temperature interacting and non-interacting results shows a growing deviation as we increase $\mu$.
Already, for the first charge jump a significant change is deduced suggesting a noticeable influence from the interactions.
Furthermore, as we will see in section~\ref{sub-sec:spectrum}, the 4-fold degeneracy around $E\sim 1$ splits, and we expect the jump of $\Delta Q = 8$ to break into jumps of size $\Delta Q = 2$.
A final assessment on the importance of the interaction in this molecule, however, cannot be made, as only one non-physical, interaction value is considered.

Furthermore, since a typical level of doping is expected to be only a few elemental charges~\cite{gregg2001doping,jacobs2017controlling}, we argue that the NLO-plane HMC provides an acceptable signal at values of $\mu$ in the relevant range for perylene.

\subsection{Extracting Energies}\label{sub-sec:extractingEnergies}
Each ensemble, fixed by a choice for $\Nt$, $\beta$, and $\mu$, results in 20 correlators; a total of $1980$ correlators need to be analyzed.
Using the fitting routine described in Appendix~\ref{apx-sec:analysis_details}, we perform about 30 to 100 fits (depending on the fit intervals and the minimum of the correlator) with either two or three exponential terms in the model for the central value and for each of the $\Nbst$ bootstrap samples.
With $\Nbst=\num{500}$ this results in $\order{10^8}$ fits.
This sheer number emphasizes that an automatic fitting procedure with well formulated criteria is needed.
In this section we discuss a selection of correlators and how their corresponding energies are extracted.
We focus in particular on the finest lattice spacing ($\Nt = 96$) and the lowest temperature ($\beta = 8$).

As discussed in Appendix~\ref{subsec:effective masses} the single particle spectrum contains positive and negative energies and the spectral decomposition can be split into increasing and decreasing exponentials.
This motivates the fit model 
\begin{equation}
\begin{aligned}
    C_{\Lambda_i}(\tau)
    &= 
    z_0^L
    e^{- E_0^{L} \tau}
    +
    z_0^R
    e^{ E_0^{R} (\tau-\beta)} \\
    &+ \sum_{n>0}^{N_L}
    z_n^L
    e^{- \left(\Delta E_n^{L} + E_{n-1}^{L} \right) \tau} \\
    &+
    \sum_{n>0}^{N_R}
    z_n^R
    e^{ \left(\Delta E_n^{R} + E_{n-1}^{R}\right) (\tau-\beta)}
\end{aligned}
    \label{eq:exp-model}
\end{equation}
where the $L$ and $R$ labels indicate whether the contribution is large at small or large $\tau$ and we have dropped the state label on the fit parameters.
Notice that the parameters $E_n^{L/R}$ and the respective splittings $\Delta E_n^{L/R} = E_{n}^{L/R} - E_{n-1}^{L/R}$ are positive.
Thus accessing the desired energy requires us to identify the dominant contribution and assign
\begin{equation}
    E^{\Lambda_i}_{U=2} = E_0^L \text{ or } -E_0^R\,.
\end{equation}
For more details please refer to appendix~\ref{subsec:Fitting}.

\subsubsection{$\mu = 0$}

The smallest energy, in magnitude, is most interesting as it moves across zero for finite chemical potential first, indicating the previously discussed charge jump. 
These energies come from the state $B_3^3$ (negative energy) and $B_1^3$ (positive energy).

\begin{figure}
    \resizebox{\linewidth}{!}{\includegraphics{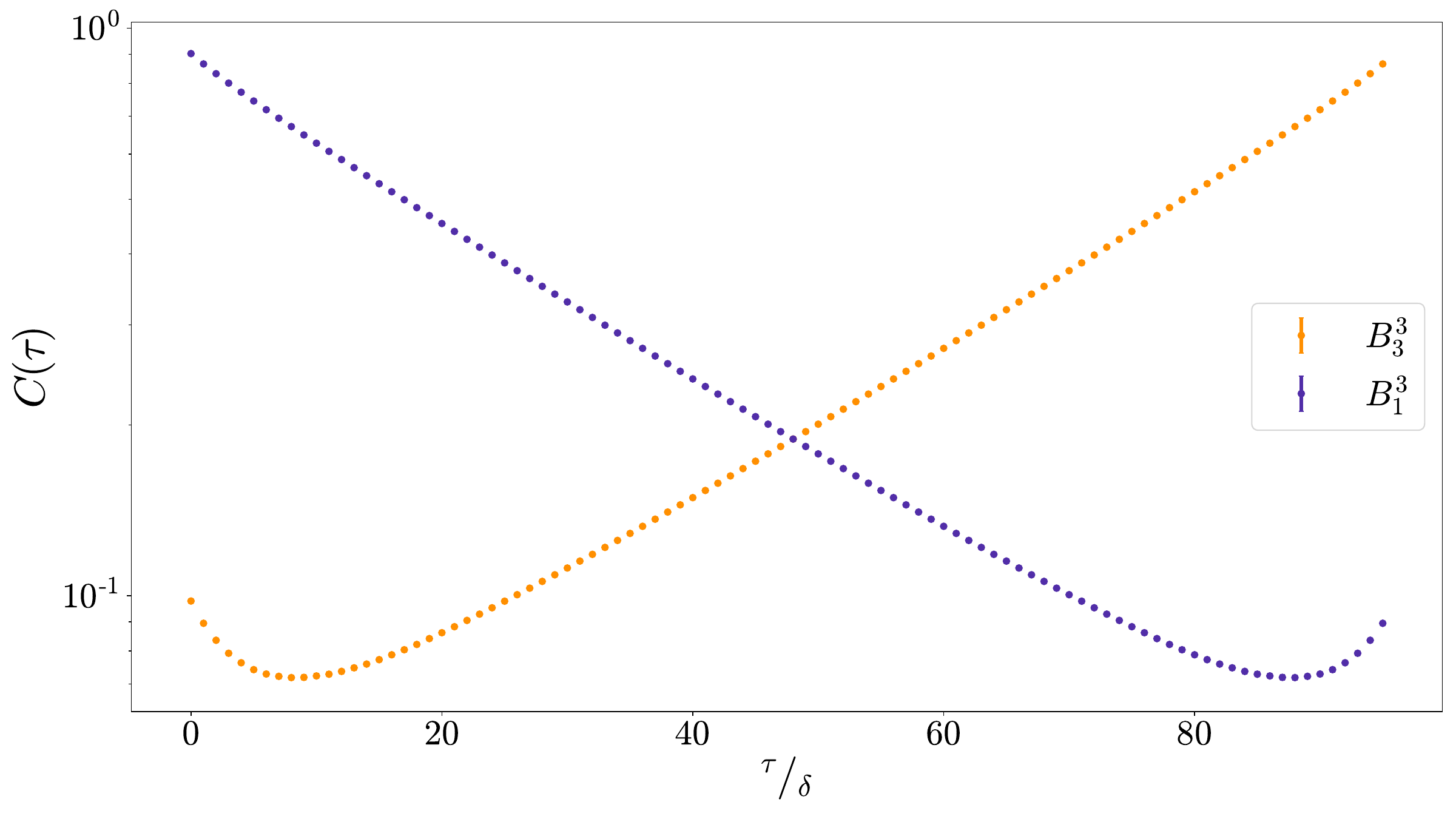}}
    \caption{ 
        Correlators corresponding to states $B_3^3$ \& $B_1^3$, orange and blue respectively.
        These are estimated at $\Nt=96, \, \beta = 8, \, U = 2, \, \mu = 0$ and correspond to the smallest, in magnitude, energy of the system.
        Uncertainties are less then 1\% making them hard to spot.
    }\label{fig:smallEnergyCorr}
\end{figure}

The $B_3^3$ and $B_1^3$ correlators at $\mu=0$ are displayed in figure~\ref{fig:smallEnergyCorr}.
The uncertainties at each time point are less then $1\%$ which results from relatively high statistics $\sigma \sim \order{\nicefrac{1}{\sqrt{10000}}}$ and the fact that its decay is relatively mild.
Especially for larger energy correlators we find a signal-to-noise problem around the minimal point.
As the energies of the $B_3^3$ and $B_1^3$ correlators differ only in sign, we find them equal up to time reversal. 
Furthermore, on a log scale they appear extremely straight for a large range of euclidean time $\tau$ indicating little excited state contamination.

\begin{figure}
    \resizebox{\linewidth}{!}{\includegraphics{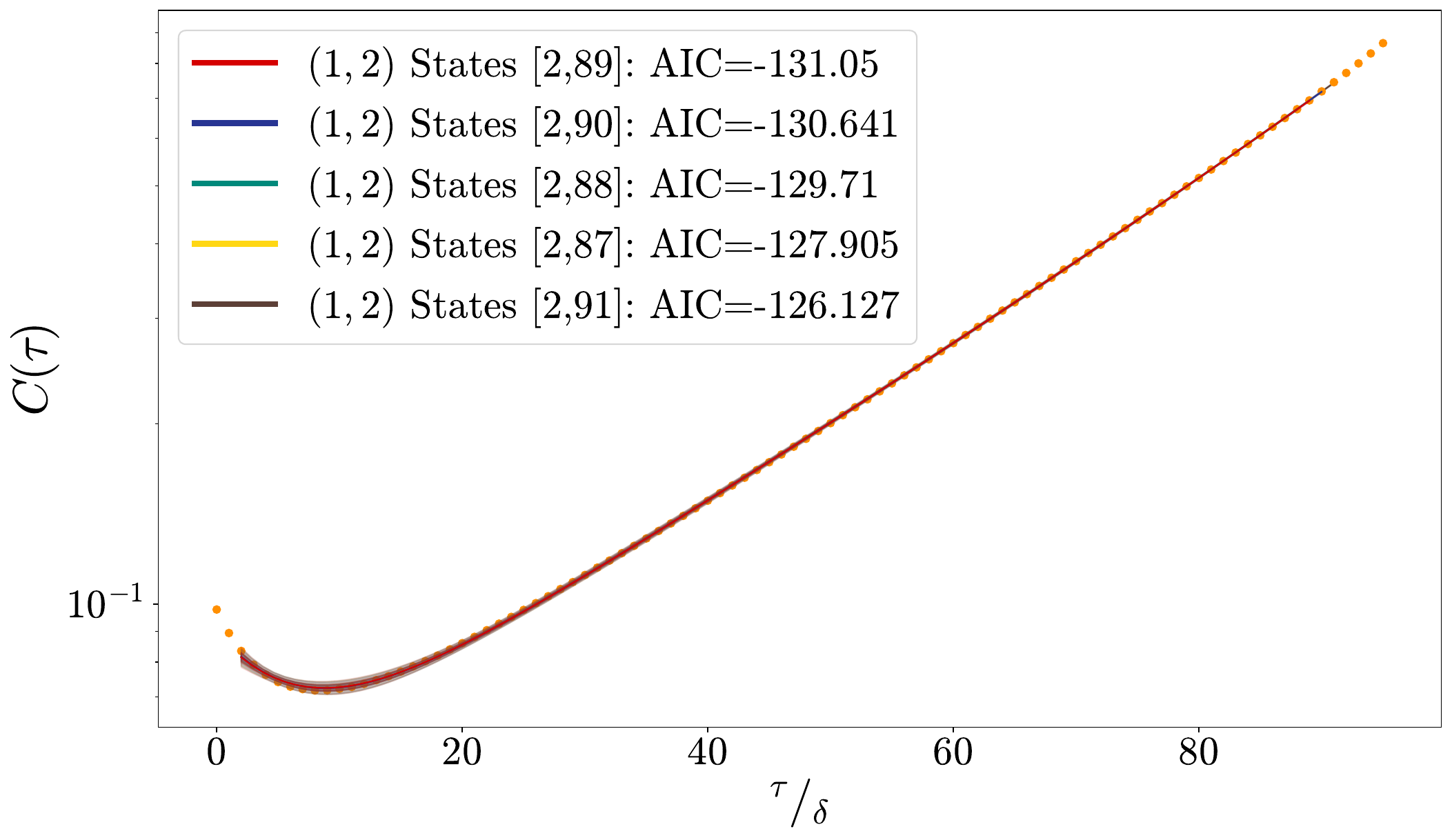}}
    \caption{ 
        Best $B_3^3$ fits. 
        The data points are the same as in figure~\ref{fig:smallEnergyCorr}. 
        For each fit, two confidence bands are plotted corresponding to one and two $\sigma$.
        These best fits are performed according to the model~\eqref{eq:exp-model} with $(N_L=1,N_R=2)$.
        No $N_R=1$ fit is in the best five.
        The fit range is indicated in the square brackets expressing values of $\nicefrac{\tau}{\delta}$ comprising almost the entire correlator.
    }\label{fig:smallEnergyCorr_bestFits}
\end{figure}

In figure~\ref{fig:smallEnergyCorr_bestFits} the 5 best fits are plotted on top of the correlator.
The data points represent the correlator, the solid lines are the fits colour-coded as indicated in the legend, and the bands indicate the one- and two-$\sigma$ confidence interval on the fit.
All these fits have two exponentials on the right while the left side has one exponential.
We also performed fits with only one exponential on the right but none are among the 5 best fits shown here.
Visually all these fits are extremely close to the data points; quantitatively the $\nicefrac{\chi^2}{\mathrm{dof}} \sim 1$ as desired for good fits.
Appendix~\ref{subsec:Model Averaging} explains how we model average fits.
Furthermore, the best fit resulting over a fitting range of $\tau/\delta \in [2,89]$ with $\nicefrac{\chi^2}{\mathrm{dof}} = 0.53$.
Its result is displayed in table~\ref{tab:bestFit_mu0}.
From here we see that excited states are clearly distinguished providing additional evidence for a reliable estimate.
\begin{table}[h!]
\begin{tabular}{lcl|lcl}
    \toprule
    \toprule
    $E_0^L$       &=&$1.749(63)$  & $z_0^L$&=&$0.04277(84)$ \\
    $E_0^R$       &=&$0.3228(93)$ & $z_0^R$&=&$0.590(56)$   \\
    $\Delta E_1^R$&=&$0.270(29)$  & $z_1^R$&=&$0.290(55)$.   \\
    \bottomrule
    \bottomrule
\end{tabular}
\caption{
    Best fit results of the $B_3^3$ correlator displayed in figure~\ref{fig:smallEnergyCorr_bestFits}.
    Uncertainties are determined through bootstrap while central values come from a fit to the central values of the data.
}\label{tab:bestFit_mu0}
\end{table}

\begin{figure*}
    \resizebox{0.49\linewidth}{!}{\includegraphics{ 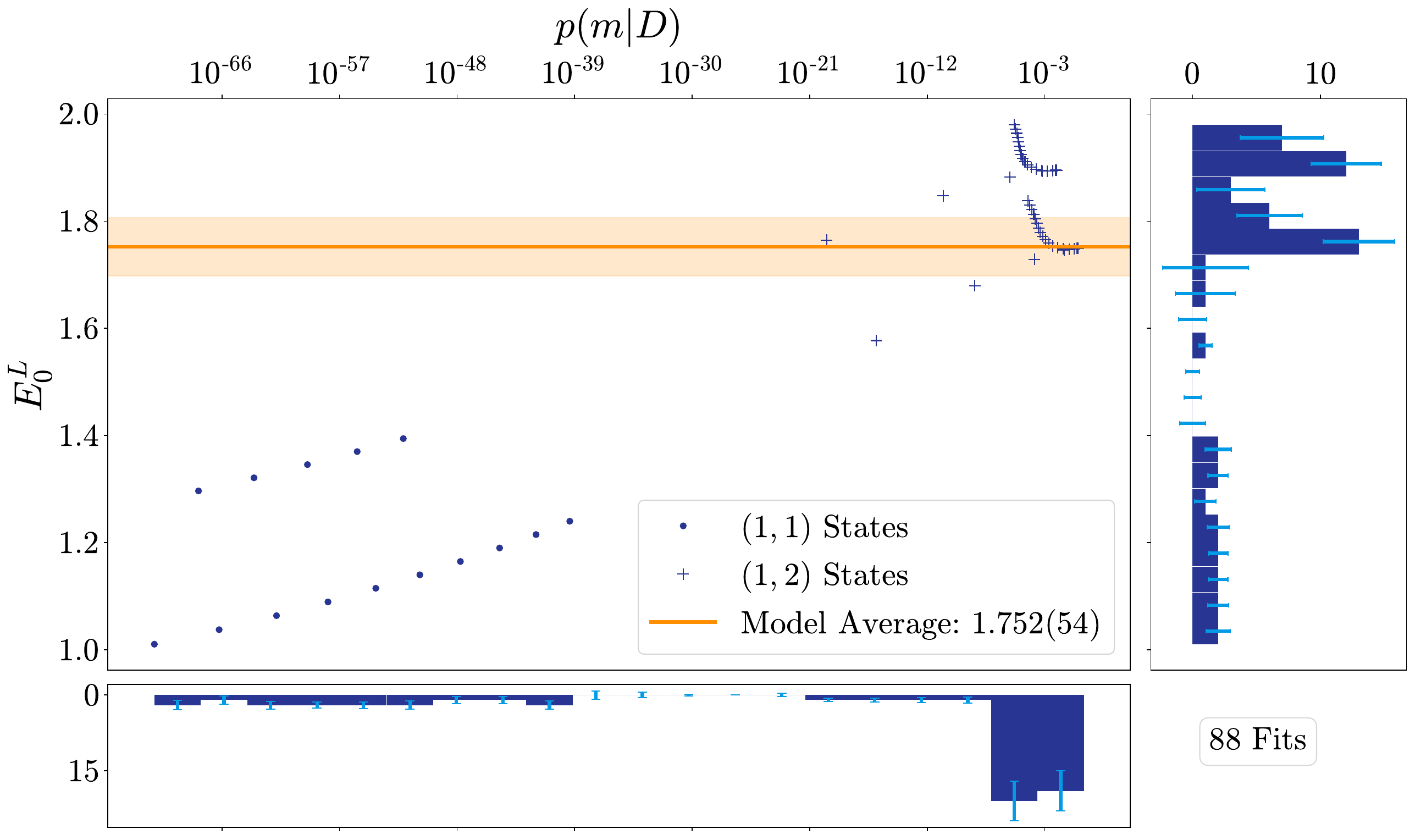}}\hfill
    \resizebox{0.49\linewidth}{!}{\includegraphics{ 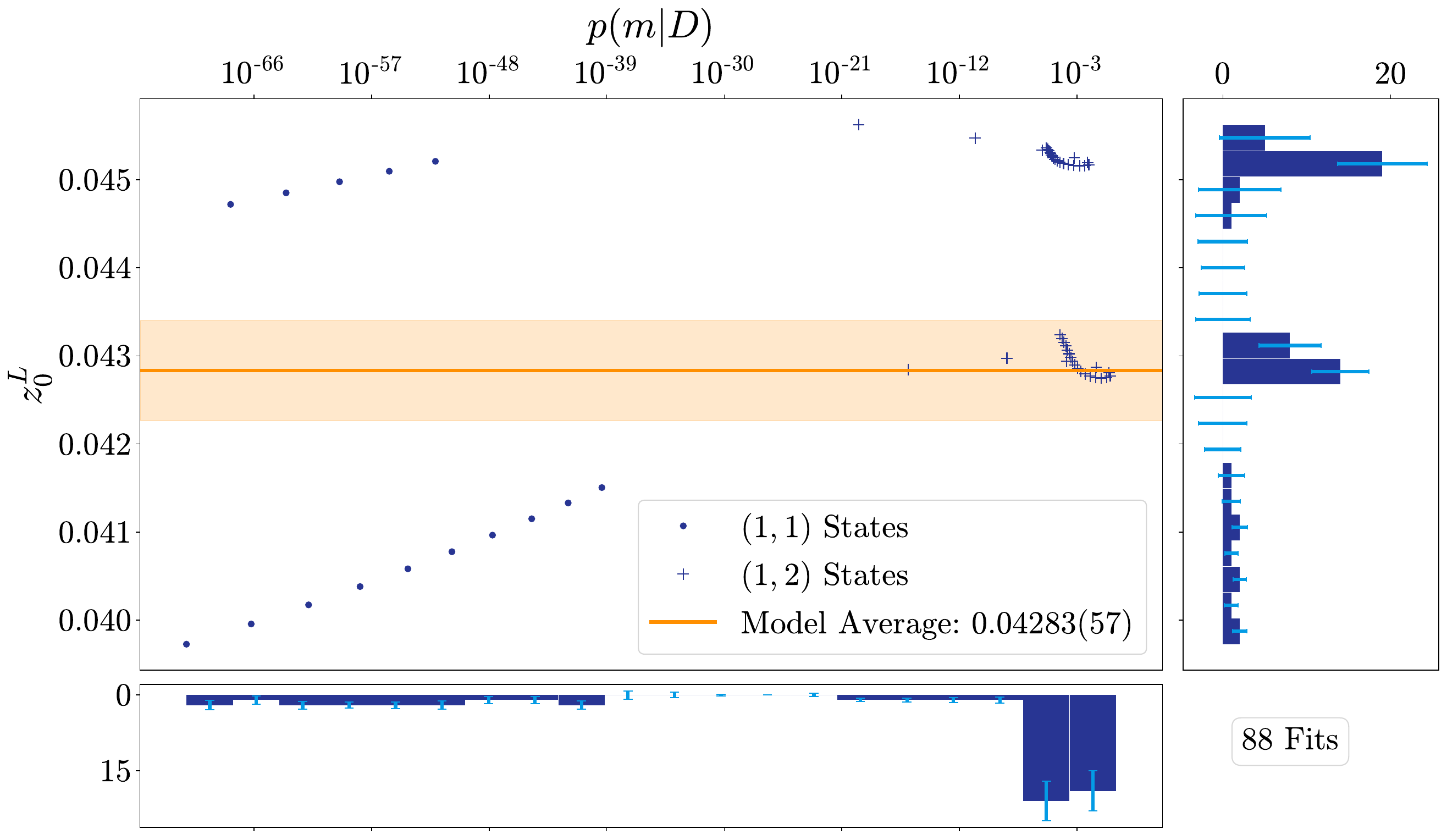}}\hfill
    \resizebox{0.49\linewidth}{!}{\includegraphics{ 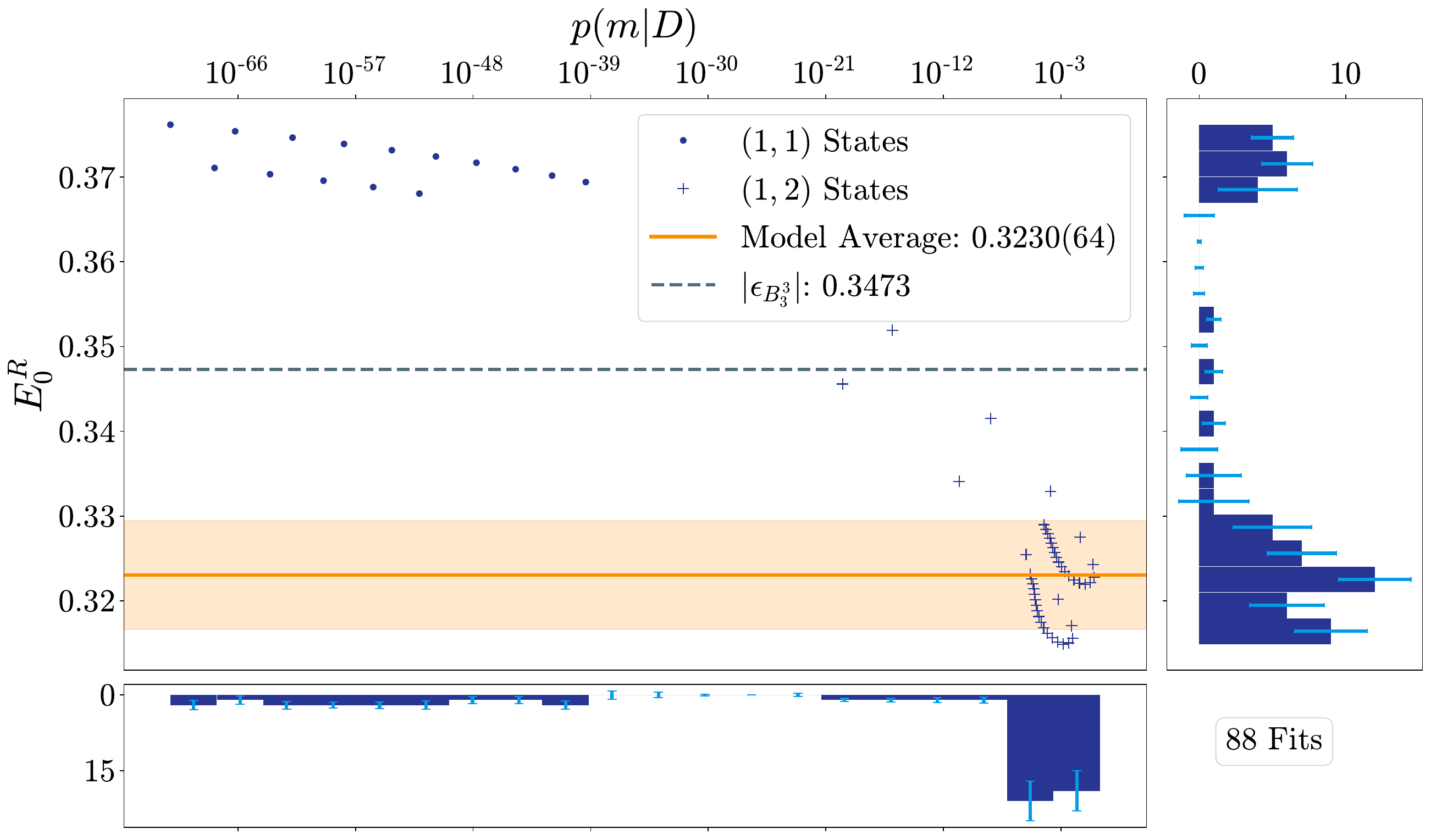}}\hfill
    \resizebox{0.49\linewidth}{!}{\includegraphics{ 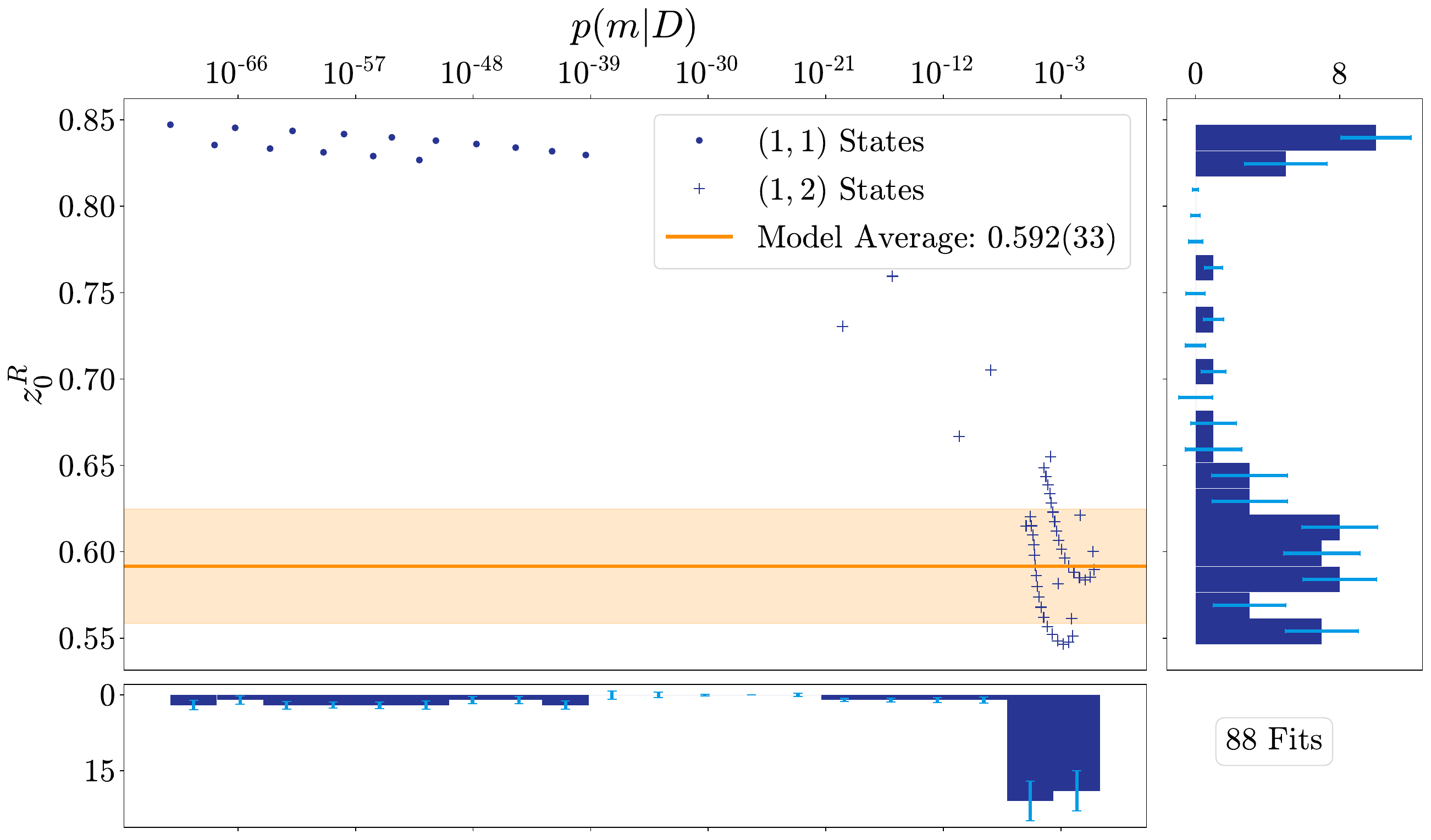}}\hfill
    \resizebox{0.49\linewidth}{!}{\includegraphics{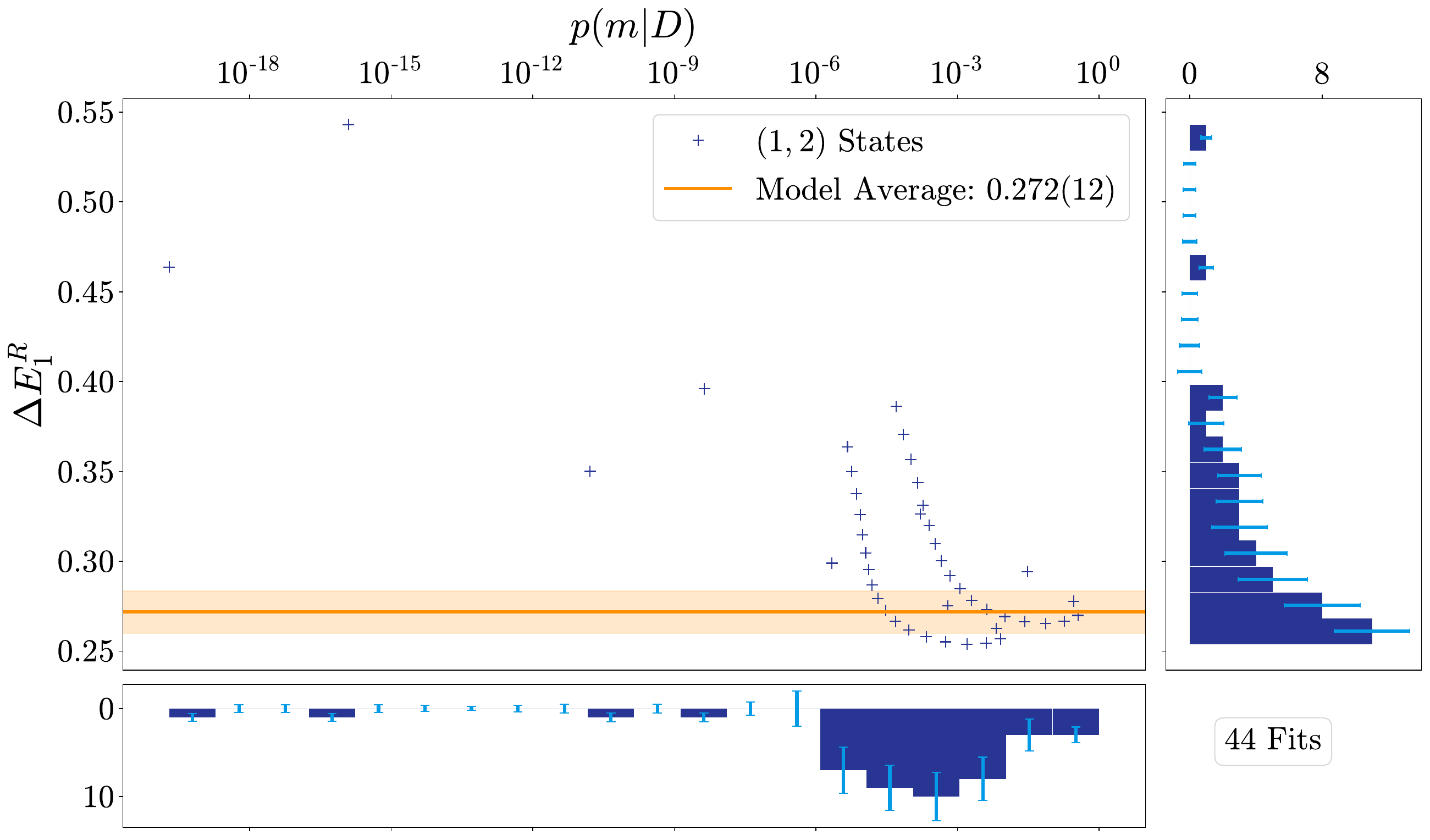}}\hfill
    \resizebox{0.49\linewidth}{!}{\includegraphics{ 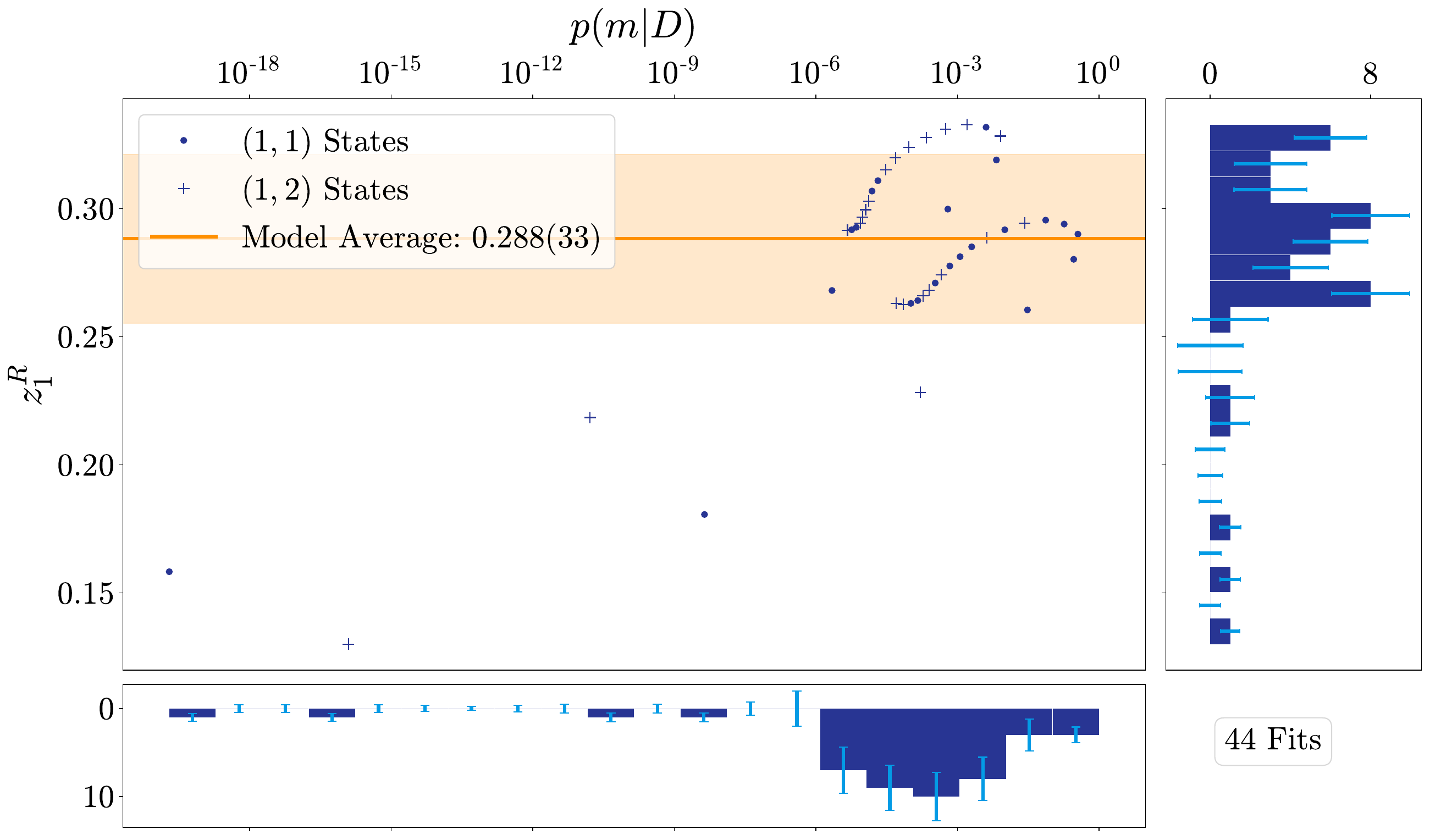}}\hfill
    \caption{
        Fit overview plots for a fit to the $B_3^3$ correlator at $\Nt=96, \, \beta = 8, \, U = 2, \, \mu = 0$. 
        A subfigure is dedicated for each parameter in the fit model~\eqref{eq:exp-model} as a function of the model weight $p(m\vert D)\sim \exp(-\nicefrac{1}{2} \mathrm{AIC})$.
        One-state fits, $(N_L = 1, N_R = 1)$, are indicated with circles,
        while two-state fits, $(N_L=1,N_R=2)$, are plotted with pluses. 
        The model average is indicated through a solid line with adjacent uncertainty determined by 
        the standard deviation of the model average on each bootstrap sample.
        Attached to the axes are counts of the fit results (unweighted) and the model weights.
        The correlation of the mode of the fit results with the mode of the weights indicate the support for the two-state fits.
        Uncertainties are only displayed on the counts, computed by bootstrapping the heights on fixed bin widths.
        Fits with $\mathrm{AIC}>200$ are not shown.
        Finally the total number of fits is shown in the lower right corner, with less fits for parameters only available in the two state fits.
    }\label{fig:smallEnergyCorr_overview}
\end{figure*}

We can further assess the stability of the fitting procedure by considering the overview plots in figure~\ref{fig:smallEnergyCorr_overview}.
The main body for each figure shows the value of the fit parameter as a function of the model probability given the data~\cite{jay2021bayesian,neil2024improved,neil2023model}
\begin{equation}
    p(m\vert D) \sim e^{-\frac{1}{2} \AIC},
\end{equation}
where AIC is the Akaike information criterion, as explained in Appendix~\ref{subsec:Model Averaging}.
A model $m$ is defined by the number of exponentials $(N_L,N_R)$ in the fit function~\eqref{eq:exp-model} and the range of euclidean time it is evaluated on.
In figure~\ref{fig:smallEnergyCorr_overview} the $(1,1)$ and $(1,2)$ state fits are plotted as circles and pluses, respectively.
These points represent the central value fit, uncertainties are not drawn.
We find the correlator to be predominantly increasing, resulting in the choice of varying $N_R$ and identifying the lowest energy to be negative.
We highlight the $p(m\vert D)$-weighted (model) average,~\eqref{eq:model-average}, of each parameter with a solid line and the uncertainty as a band. 
This uncertainty is obtained by the standard deviation of the model average over all bootstrap samples.
For the $E_0^R$ the absolute value of the non-interacting energy is added as a grey dashed line to provide a reference.
Attached to the ordinate and abscissa are the counts of the parameters and model weights (histograms).
They visualize the distribution of the fit results.
The total number of fits done is indicated in the lower right corner. 
This number is naturally smaller for the parameters only appearing in the two state fits.

Overall, we find great stability in these fits, 
as evidenced by the string of points converging towards larger weights. 
The two bands in all figures originate from the two allowed fit interval starting points at $\nicefrac{\tau}{\delta} = 1, 2$ for the fits.
As the AIC penalizes additional parameters, we find significant support for the $(1,2)$ state fits;
their respective mode is strongly correlated with the mode of the weights.
The fact that the best fits almost span over the entire abscissa strengthens this even further.

Finally, this fitting procedure results in the model averaged energy
\begin{equation}
    E^{B_3^3}_{U=2}\left(\Nt = 96, \beta = 8 \, \vert \, \mu = 0 \right) = -0.3230(64).
\end{equation}

\subsubsection{Transition of the Smallest Energy at $\mu\ne0$}
\begin{figure}
    \resizebox{\linewidth}{!}{\includegraphics{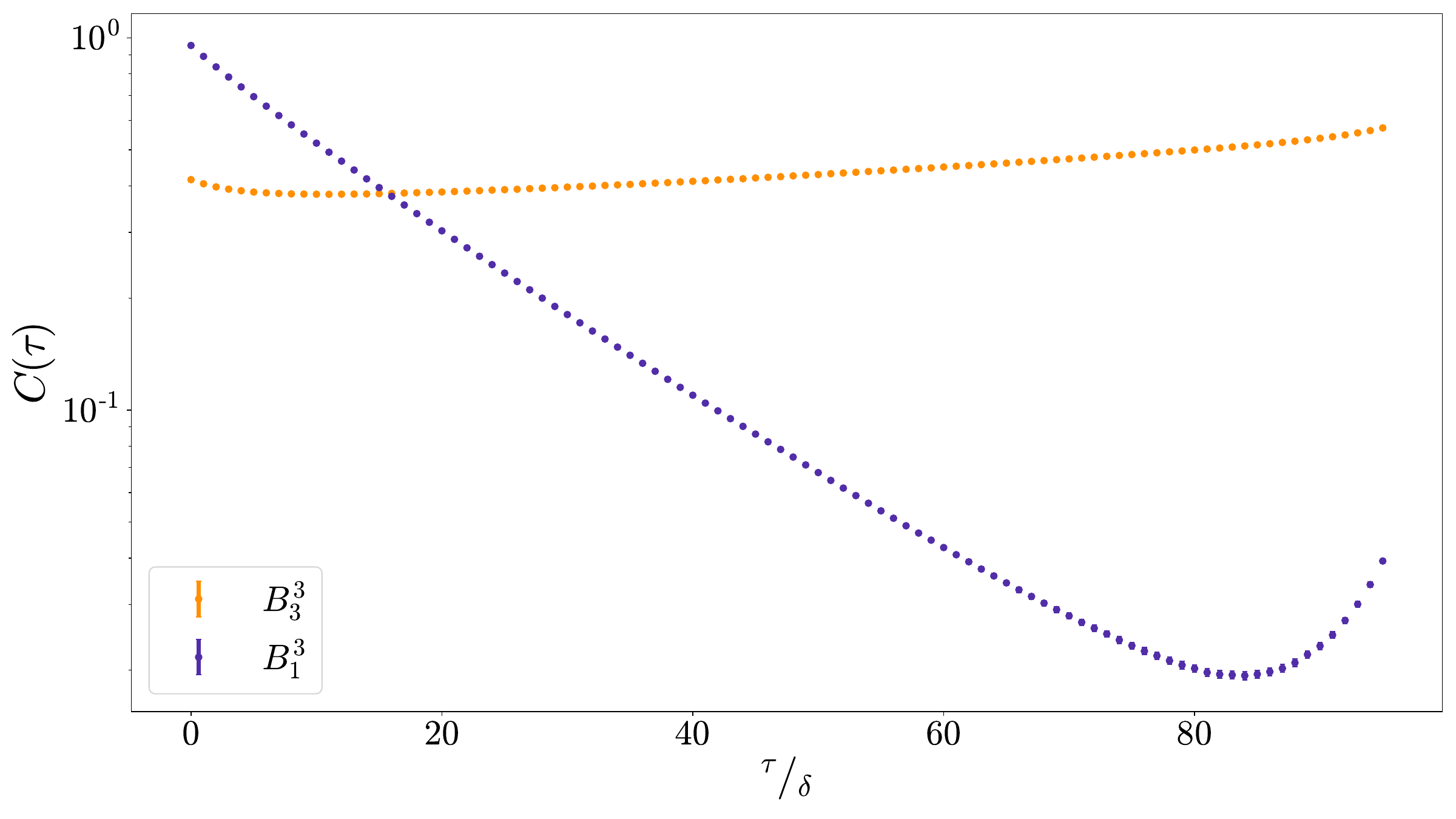}}\hfill
    \resizebox{\linewidth}{!}{\includegraphics{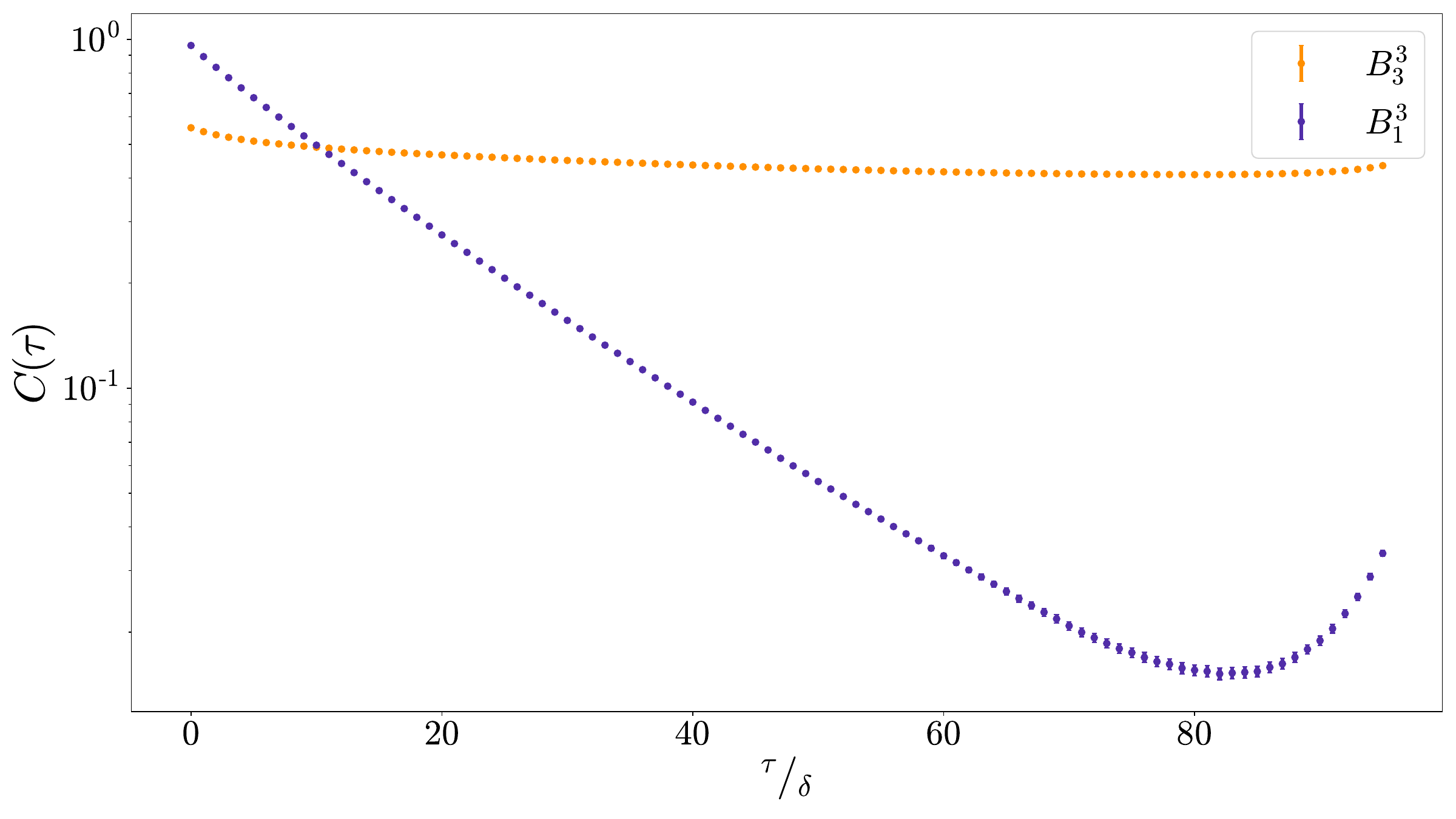}}\hfill
    \caption{ 
        Correlators corresponding to states $B_3^3$ (orange) and $B_1^3$ (blue) estimated with $\Nt=96, \, \beta = 8, \, U = 2$.
        While figure~\ref{fig:smallEnergyCorr} shows $\mu=0$, the upper panel shows $\mu = 0.4$ and the lower panel $\mu = 0.5$.
        The transition of the smallest negative energy to a positive energy happens in between these two values.
        The trend of both correlators is toward a more negative slope with increasing chemical potential, as expected given our sign convention for $\mu$.
    }\label{fig:smallEnergyCorr_finMu}
\end{figure}

The interacting energies change with $\mu$, and sometimes a state's energy changes sign.
This happens, for instance, between $\mu = 0.4$ and $0.5$, where the $B_3^3$ state's energy crosses 0.
To illustrate the effect of the chemical potential on the correlator, figure~\ref{fig:smallEnergyCorr_finMu} again shows the two states $B_3^3$ and $B^3_1$; between $\mu=0.4$ (the upper panel) and $\mu=0.5$ (lower panel) the $B_3^3$ correlator goes from predominantly decreasing to predominantly increasing, indicating an energy crossing 0.
Even at these chemical potentials we find a great resolution accounting for a statistical power of 
$\abs{\expval{\Sigma}} = 0.6228(46), 0.4707(59)$ at $\mu=0.4,0.5$ respectively.
We emphasize that without alleviating the sign problem with a contour deformation these correlators are overwhelmed by noise and no results can be extracted.

\begin{figure}
    \resizebox{\linewidth}{!}{\includegraphics{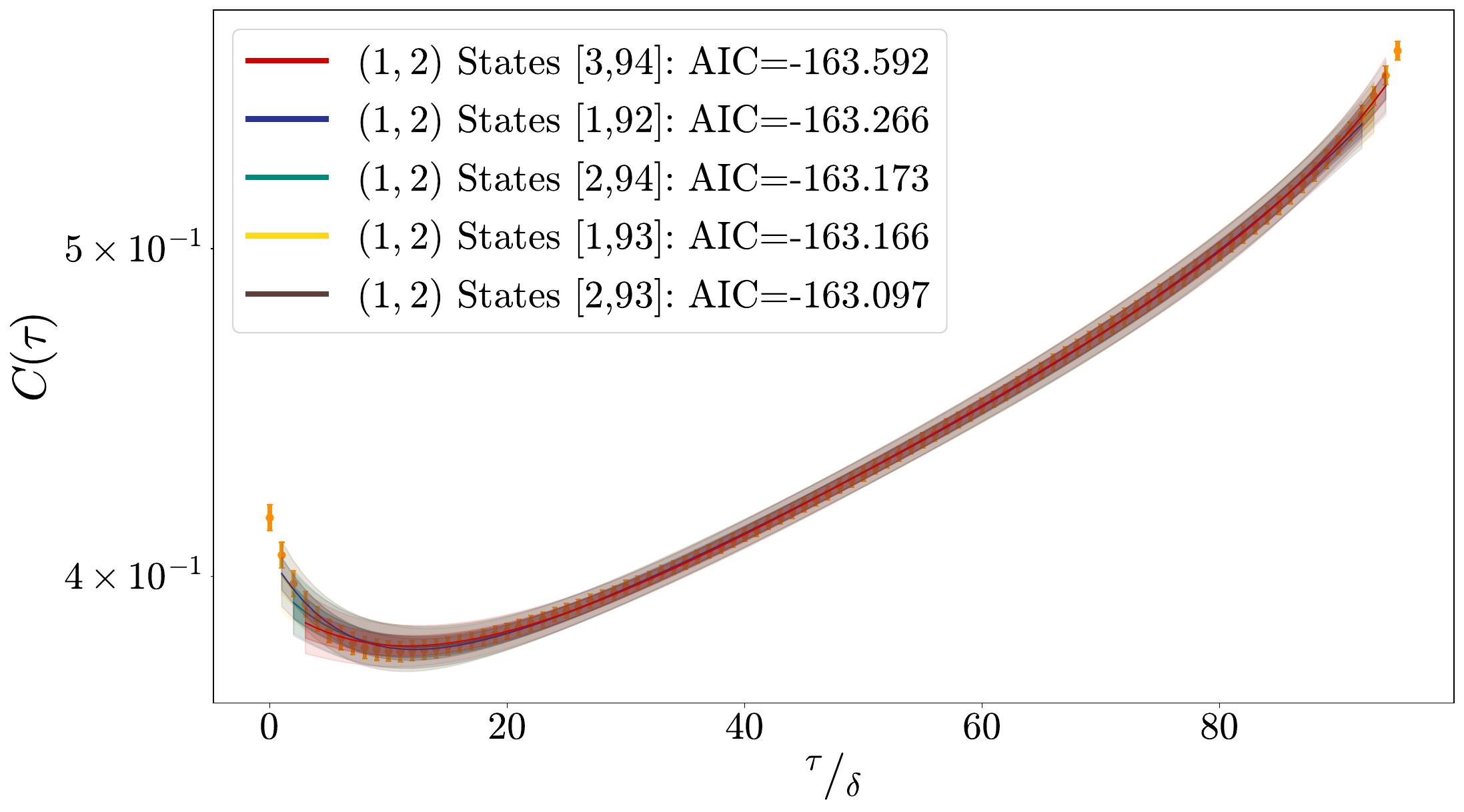}}\hfill
    \resizebox{\linewidth}{!}{\includegraphics{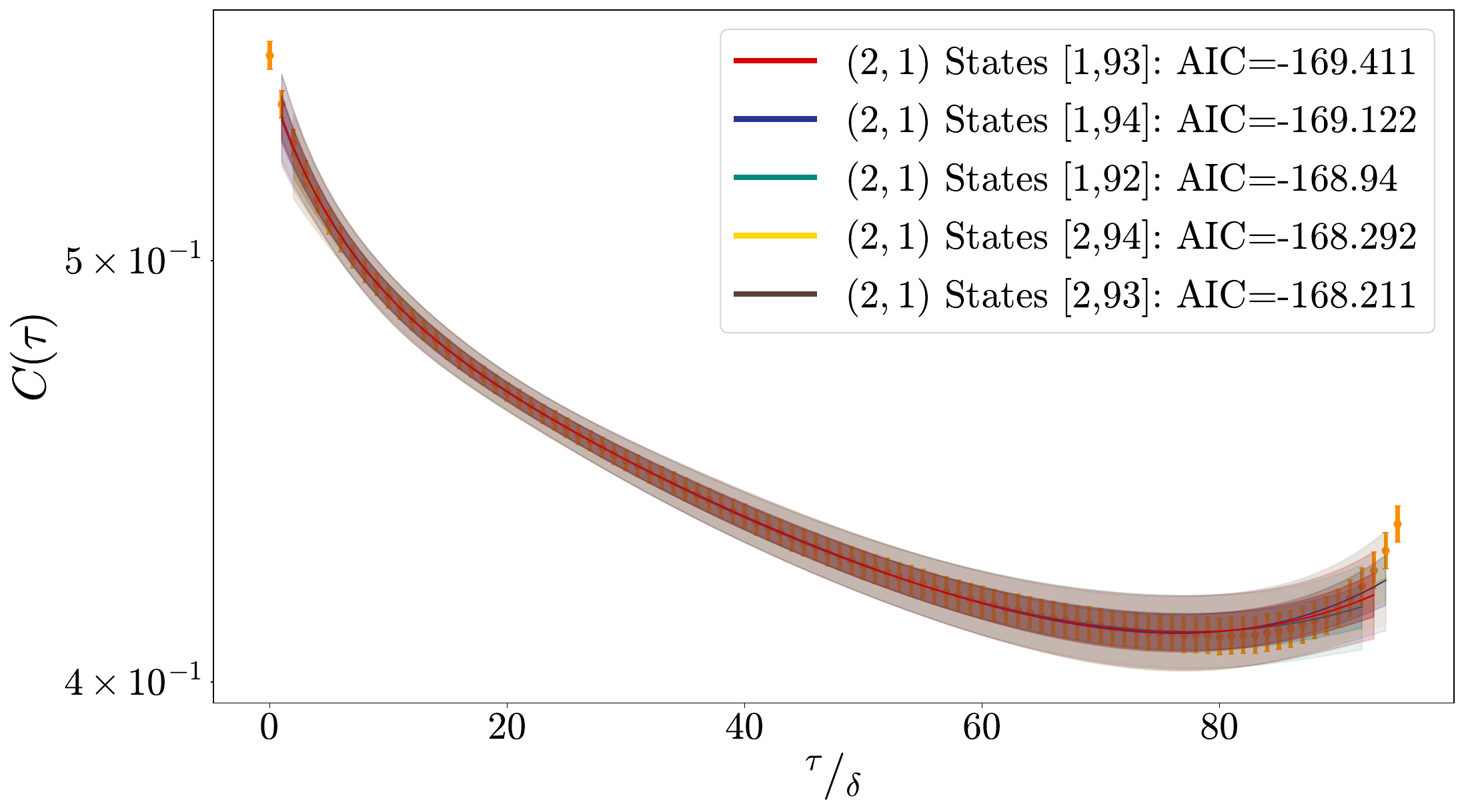}}\hfill
    \caption{ 
        Best $B_3^3$ Fits. 
        The data points are the same as in figure~\ref{fig:smallEnergyCorr_finMu}, at again $\mu=0.4$ (upper) and $\mu = 0.5$ (lower). 
        Similarly, each fit is plotted as a solid line with two confidence bands corresponding to one and two $\sigma$.
        Further, the best fits are again with two states $(N_L=1,N_R=2)$ and $(N_L=2,N_R=1)$ respectively.
        The fit range is indicated in the square brackets expressing values of $\nicefrac{\tau}{\delta}$ comprising almost the entire correlator.
        The transition from negative to positive energy can be seen better on this scale.
    }\label{fig:smallEnergyCorr_bestFits_finMu}
\end{figure}

In figure~\ref{fig:smallEnergyCorr_bestFits_finMu}, similar plots for the 5 best fits to the $B_3^3$ correlator are shown.
As in figure~\ref{fig:smallEnergyCorr_finMu}, the upper panel shows $\mu = 0.4$, while the lower panel shows $\mu= 0.5$.
The best of these fits have $\nicefrac{\chi^2}{\mathrm{dof}} = 0.11, 0.075$.
The overview plots exhibit the same nice features as in the $\mu = 0$ case and we omit them for concision.

For these two fits we find model average energies, 
\begin{align}
    E^{B_3^3}_{U=2}\left(\Nt = 96, \beta = 8 \, \vert \, \mu = 0.4 \right) &= -0.0480(37), \\
    E^{B_3^3}_{U=2}\left(\Nt = 96, \beta = 8 \, \vert \, \mu = 0.5 \right) &= +0.0427(73).
\end{align}
 
\subsection{Continuum Limit}\label{sub-sec:contLimit}
\begin{figure*}
    \resizebox{!}{0.9\textheight}{\includegraphics{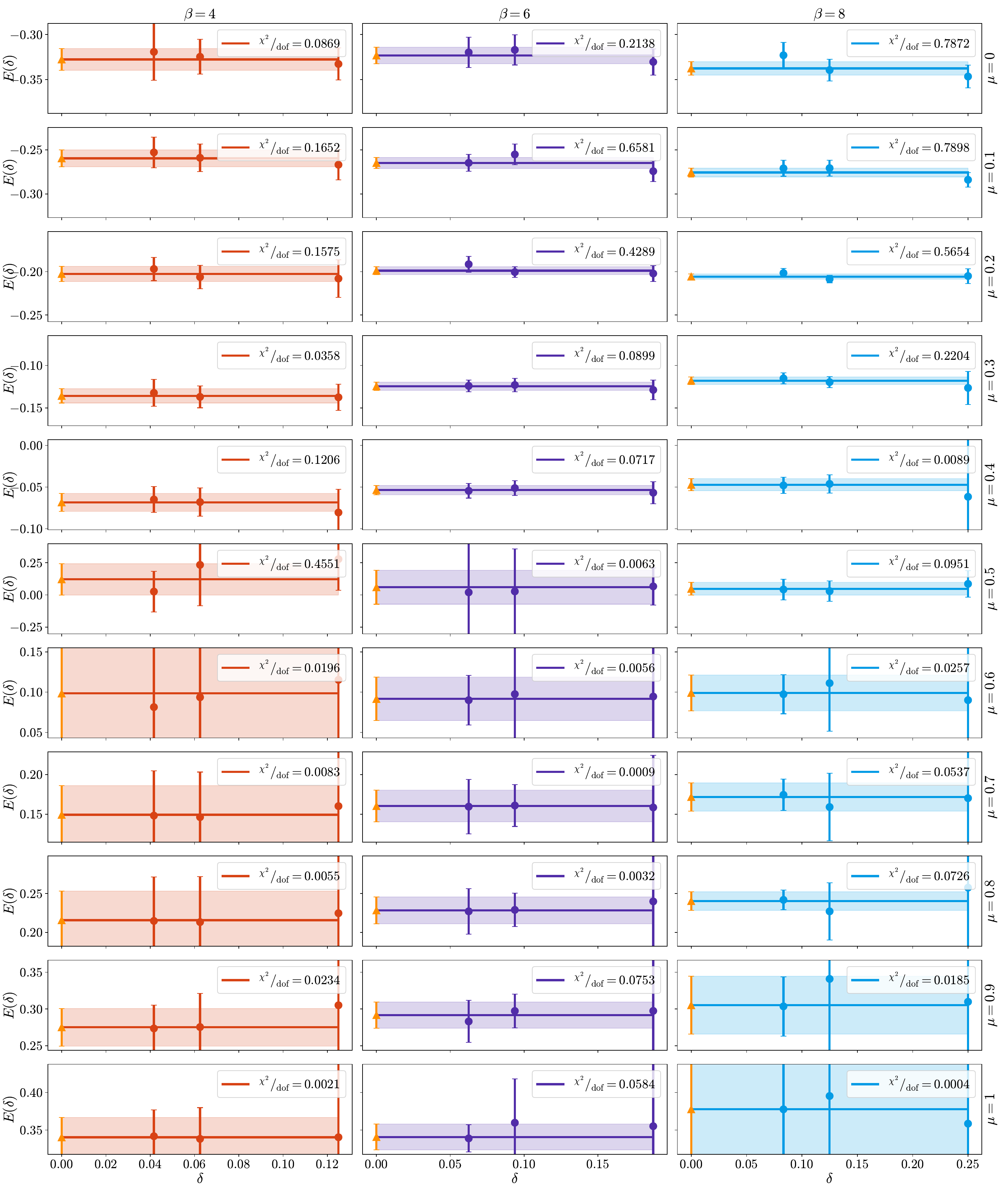}}
    \caption{ 
        Continuum limit of the $B_3^3$ as a function of chemical potential (top to bottom) for the three available temperatures left to right. 
        The scale of the ordinate is kept to provide a feeling on the $\beta$ dependence. 
        The title of each subplot shows the validity via a $\nicefrac{\chi^{2}}{\mathrm{dof}}$. 
        As these are just fits to a constant, smaller values stem from uncertainties thus all fits are very good.
    }\label{fig:continuumEnergy}
\end{figure*}

To remove the systematic errors introduced by discretizing the thermal trace we must perform a continuum limit $\delta \to 0$.
Given our data, at each $\beta$ we can fit a constant, as shown for the $B_3^3$ state in figure~\ref{fig:continuumEnergy}.
The inverse temperature $\beta$ increases across the columns and the chemical potential increases down the rows.
In each row the ordinate maintains the same scale to provide a rough idea of the $\beta$ dependence.
A triangle at $\delta = 0$ indicates the continuum value; a corresponding solid line is put to guide the eye toward larger $\delta$.

The legend gives the $\nicefrac{\chi^2}{\mathrm{dof}}$ for the constant fit.
All states give values between $\nicefrac{\chi^2}{\mathrm{dof}} = \num{3.5e-4}$ and $\num{0.8}$.
Overall, the residuals are significantly smaller than one would expect for an ideal fit $\nicefrac{\chi^2}{\mathrm{dof}}\approx 1$.
In particular the very small $\chi^2$-values are governed by the increased uncertainties at larger chemical potentials.
At this point we want to emphasize that the purely statistical uncertainties on the best fits are significantly smaller. 
However, due to the bootstrap over model averages we include systematics, from the choice of fit model, in the uncertainties.
This conservative error estimation allows us to be very confident about the correctness of our results within the provided uncertainty range.

We find that all ensemble's extrapolations are extremely flat, showing little dependence on the lattice spacing at the chosen parameters.
Linear contributions are not well-supported by the data, see Appendix~\ref{subsec:Limits} for a thorough discussion.
We discard the spectrum at $\mu = 1.1$ as the noise is too large to extract the higher energies for $\beta=8$ reliably without additional samples.

\subsection{Spectrum}\label{sub-sec:spectrum}
\begin{figure*}
    \centering
    \resizebox{!}{0.9\textheight}{\includegraphics{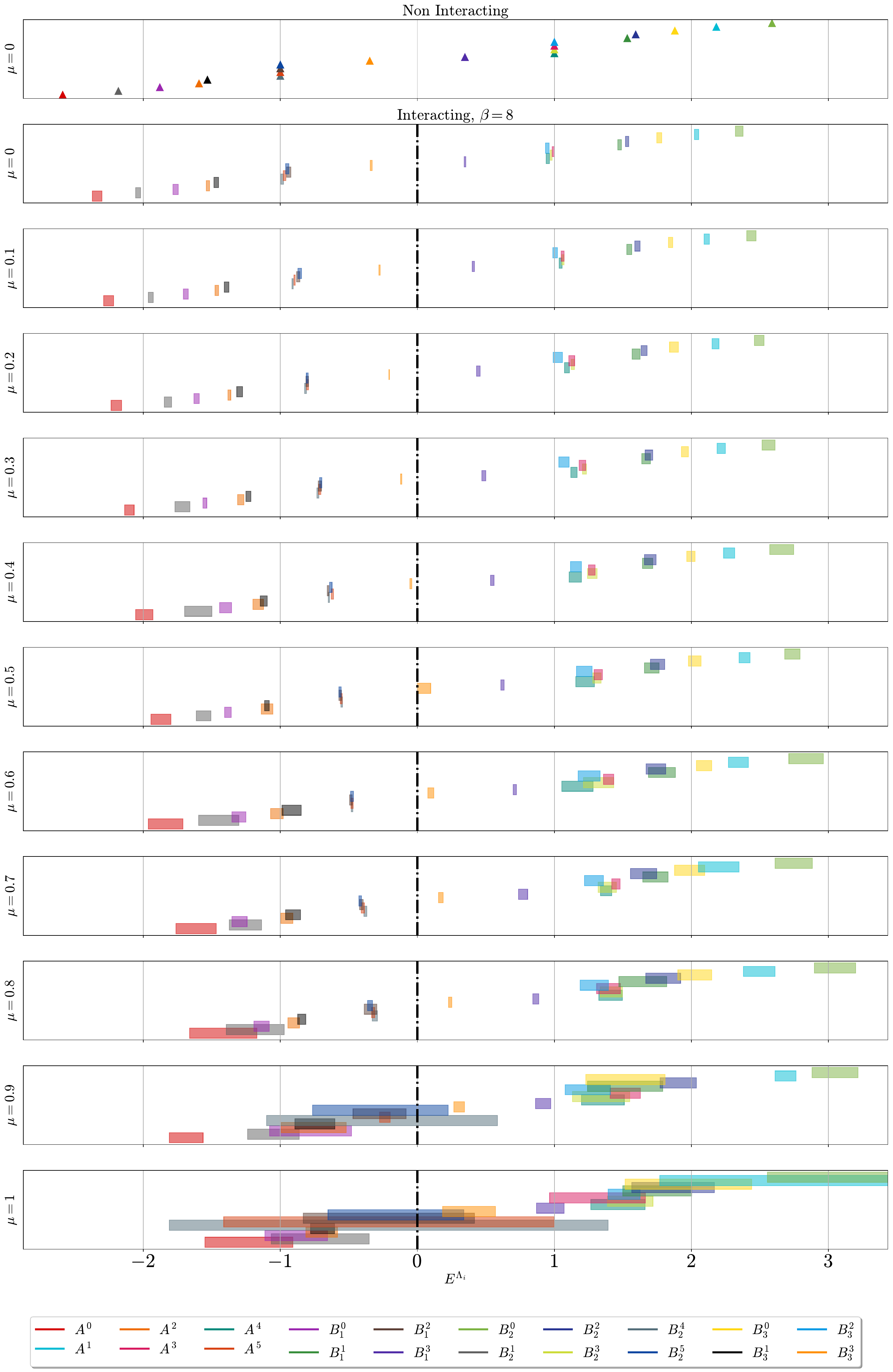}}
    \caption{
        Spectrum plot at $\beta = 8$ shows the single particle spectrum as a function of chemical potential.
        The first panel indicates the non-interacting result at $\mu = 0$, all subsequent panels are at finite interaction, $U = 2$.
        Notably, the accidental degeneracy at $E_{U=0} = \pm 1$ is split into eight energy levels.
        Following the chemical potential, $B_3^3$ can be observed moving towards zero energy and transitioning out of the fermi sea past $\mu = 0.4$ as expected from the total system charge discussed in figure~\ref{fig:charge-per-mu}.
        A second transition is not resolved at available statistics.
    }\label{fig:spectrum}
\end{figure*}

We can now collect all continuum energies and plot them as a function of chemical potential.
We present this result in two ways, first with the barcode plot in figure~\ref{fig:spectrum} that provides an overview on how the spectrum behaves as function of $\mu$.
Each panel in this figure details 20 single-particle states in $\beta=8$ spectrum at fixed chemical potential.
The first two panels offer a comparison between non-interacting and interacting spectra at $\mu=0$; each shows the expected symmetric spectrum, providing a check on the analysis.
While the small energies are very close and the ends of the spectra differ more meaningfully, we can see that the interactions split the accidental quadruplets of states at $E = \pm 1$.
Lower panels have increasing chemical potential and the energies grow with $\mu$ as expected.
In particular, the least negative state $B_3^3$ moves closer and closer to zero, changing sign after $\mu = 0.4$ as expected from the $\expval{Q}=1$ crossing in figure~\ref{fig:charge-per-mu}.
Up to $\mu =0.8$ the signal is good to resolve all energies with great precision.
Starting at $\mu = 0.9$ the sign problem becomes prevalent, providing statistical powers smaller than $\abs{\expval{\Sigma}} \leq 0.1547(75)$ resulting in significantly larger uncertainties. 
To map out the second big transition, expected after $\mu=1$ from figure~\ref{fig:charge-per-mu}, more statistics are required.
Appendix~\ref{apx-sec:moreSpectrum} details the same plots for $\beta = 4,6$ obeying a similar behaviour.

\begin{figure*}
    \centering
    \resizebox{0.825\textwidth}{!}{\includegraphics{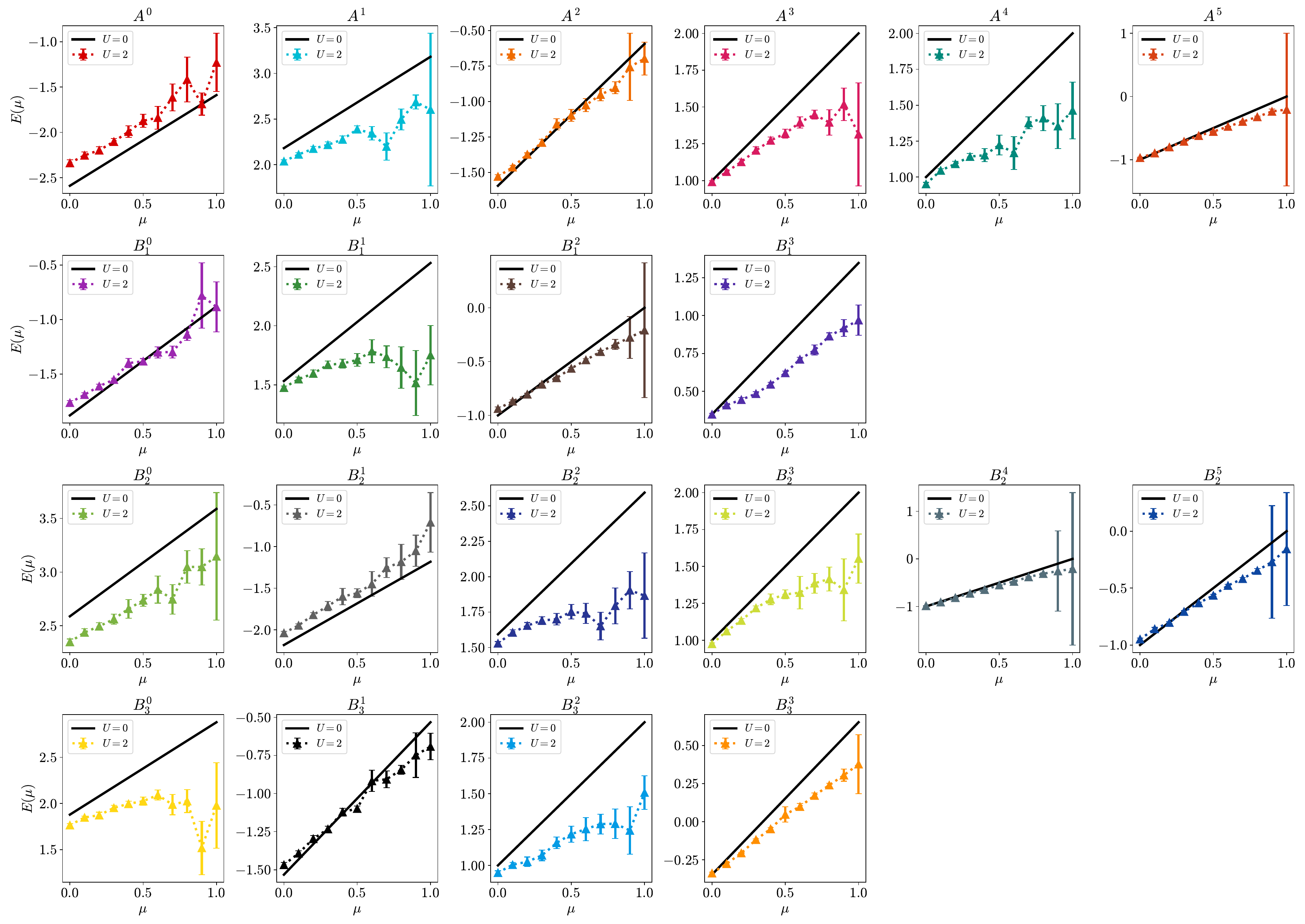}}
    \caption{
        Single particle energy overview as a function of chemical potential at $\beta = 8$.
        Each panel shows the energy of a particular state.
        The non-interacting energy is provided as a solid black line while the data points are at $U=2$.
       Difference to the non-interacting energy is more pronounced towards larger chemical potentials. 
    }\label{fig:EcontPerMu}
\end{figure*}

Second, figure~\ref{fig:EcontPerMu} details the $\mu$ dependence for each state's energy, which makes it easier to compare to the non-interacting finite-$\mu$ result.
In each panel the solid black line represents the non-interacting result, while the data points display the interacting result.
For most states a significant divergence from the non-interacting result can be seen.
As the chemical potential increases the behaviour of a given state is expected to change as the ground state changes.
Indeed we observe slightly different slopes for all states after $\mu = 0.4$. 
This is more pronounced at larger $\beta$ pointing towards a non-trivial zero temperature limit.
Finally, the energy levels at $\beta = 8$ are detailed in table~\ref{tab:spectrum}.

\begin{table*}
\begin{tabular}{l|llllllllll}
\toprule\toprule
    \multicolumn{1}{c}{$\mu$}   &  
    \multicolumn{1}{c}{$A^0$}   & 
    \multicolumn{1}{c}{$A^1$}   & 
    \multicolumn{1}{c}{$A^2$}   & 
    \multicolumn{1}{c}{$A^3$}   & 
    \multicolumn{1}{c}{$A^4$}   & 
    \multicolumn{1}{c}{$A^5$}   & 
    \multicolumn{1}{c}{$B_1^0$} & 
    \multicolumn{1}{c}{$B_1^1$} & 
    \multicolumn{1}{c}{$B_1^2$} & 
    \multicolumn{1}{c}{$B_1^3$} \\[1em]
0   & -2.337(36) & \phantom{-}2.038(16) & -1.529(12) & \phantom{-}0.9917(85) & \phantom{-}0.951(12) & -0.9696(92) & -1.763(20) & \phantom{-}1.476(14) & -0.938(16)  & \phantom{-}0.3477(68) \\
0.1 & -2.253(37) & \phantom{-}2.112(19) & -1.463(13) & \phantom{-}1.061(12)  & \phantom{-}1.046(11) & -0.8969(50) & -1.689(18) & \phantom{-}1.547(19) & -0.869(12)  & \phantom{-}0.4090(86) \\
0.2 & -2.197(38) & \phantom{-}2.176(25) & -1.373(11) & \phantom{-}1.127(21)  & \phantom{-}1.092(19) & -0.8010(69) & -1.611(20) & \phantom{-}1.596(29) & -0.8041(92) & \phantom{-}0.445(13)  \\
0.3 & -2.102(35) & \phantom{-}2.218(31) & -1.289(23) & \phantom{-}1.205(25)  & \phantom{-}1.143(23) & -0.7123(75) & -1.551(14) & \phantom{-}1.669(31) & -0.712(11)  & \phantom{-}0.484(15)  \\
0.4 & -1.992(63) & \phantom{-}2.276(41) & -1.160(39) & \phantom{-}1.273(24)  & \phantom{-}1.152(45) & -0.6201(94) & -1.400(43) & \phantom{-}1.680(38) & -0.6515(59) & \phantom{-}0.546(13)  \\
0.5 & -1.872(73) & \phantom{-}2.388(40) & -1.097(43) & \phantom{-}1.320(30)  & \phantom{-}1.224(69) & -0.5555(82) & -1.383(24) & \phantom{-}1.711(54) & -0.5635(99) & \phantom{-}0.621(13)  \\
0.6 & -1.84(13)  & \phantom{-}2.344(73) & -1.025(46) & \phantom{-}1.396(38)  & \phantom{-}1.17(11)  & -0.4762(78) & -1.301(51) & \phantom{-}1.785(98) & -0.4848(96) & \phantom{-}0.711(13)  \\
0.7 & -1.61(15)  & \phantom{-}2.20(15)  & -0.951(44) & \phantom{-}1.449(30)  & \phantom{-}1.377(41) & -0.398(12)  & -1.298(55) & \phantom{-}1.737(93) & -0.409(13)  & \phantom{-}0.772(33)  \\
0.8 & -1.42(25)  & \phantom{-}2.50(12)  & -0.903(42) & \phantom{-}1.395(87)  & \phantom{-}1.410(87) & -0.323(12)  & -1.137(56) & \phantom{-}1.65(17)  & -0.341(46)  & \phantom{-}0.864(22)  \\
0.9 & -1.69(12)  & \phantom{-}2.687(76) & -0.76(24)  & \phantom{-}1.52(11)   & \phantom{-}1.35(16)  & -0.236(39)  & -0.78(30)  & \phantom{-}1.52(27)  & -0.28(19)   & \phantom{-}0.917(56)  \\
1   & -1.23(32)  & \phantom{-}2.60(83)  & -0.70(12)  & \phantom{-}1.32(35)   & \phantom{-}1.46(20)  & -0.2(1.2)   & -0.88(23)  & \phantom{-}1.75(25)  & -0.21(62)   & \phantom{-}0.97(10)   \\
\midrule
                                &
    \multicolumn{1}{c}{$B_2^0$} & 
    \multicolumn{1}{c}{$B_2^1$} & 
    \multicolumn{1}{c}{$B_2^2$} & 
    \multicolumn{1}{c}{$B_2^3$} & 
    \multicolumn{1}{c}{$B_2^4$} & 
    \multicolumn{1}{c}{$B_2^5$} & 
    \multicolumn{1}{c}{$B_3^0$} & 
    \multicolumn{1}{c}{$B_3^1$} & 
    \multicolumn{1}{c}{$B_3^2$} & 
    \multicolumn{1}{c}{$B_3^3$} \\[1em]
0   & \phantom{-}2.350(28) & -2.037(18) & \phantom{-}1.530(12) & \phantom{-}0.9758(77) & -0.9870(96) & -0.949(12)  & \phantom{-}1.766(18) & -1.467(17) & \phantom{-}0.949(13) &          -0.3376(74) \\
0.1 & \phantom{-}2.438(34) & -1.946(18) & \phantom{-}1.606(20) & \phantom{-}1.0622(78) & -0.9099(44) & -0.859(14)  & \phantom{-}1.848(17) & -1.392(16) & \phantom{-}1.005(17) &          -0.2757(50) \\
0.2 & \phantom{-}2.496(35) & -1.819(26) & \phantom{-}1.655(22) & \phantom{-}1.134(12)  & -0.8163(65) & -0.8027(72) & \phantom{-}1.873(32) & -1.297(21) & \phantom{-}1.025(33) &          -0.2057(30) \\
0.3 & \phantom{-}2.563(47) & -1.714(55) & \phantom{-}1.691(28) & \phantom{-}1.219(15)  & -0.7259(76) & -0.7061(85) & \phantom{-}1.953(26) & -1.233(18) & \phantom{-}1.070(37) &          -0.1180(44) \\
0.4 & \phantom{-}2.658(87) & -1.599(99) & \phantom{-}1.700(42) & \phantom{-}1.277(35)  & -0.6455(53) & -0.6311(86) & \phantom{-}1.997(30) & -1.121(25) & \phantom{-}1.157(41) &          -0.0472(72) \\
0.5 & \phantom{-}2.737(56) & -1.561(53) & \phantom{-}1.753(51) & \phantom{-}1.311(30)  & -0.5514(59) & -0.5643(83) & \phantom{-}2.025(46) & -1.099(17) & \phantom{-}1.218(57) & \phantom{-}0.048(50) \\
0.6 & \phantom{-}2.84(13)  & -1.45(15)  & \phantom{-}1.740(71) & \phantom{-}1.32(11)   & -0.4763(68) & -0.477(11)  & \phantom{-}2.092(56) & -0.917(70) & \phantom{-}1.254(81) & \phantom{-}0.099(22) \\
0.7 & \phantom{-}2.75(14)  & -1.25(12)  & \phantom{-}1.651(96) & \phantom{-}1.386(67)  & -0.380(10)  & -0.417(10)  & \phantom{-}1.99(11)  & -0.906(55) & \phantom{-}1.289(69) & \phantom{-}0.172(18) \\
0.8 & \phantom{-}3.05(15)  & -1.18(21)  & \phantom{-}1.79(13)  & \phantom{-}1.415(81)  & -0.311(19)  & -0.346(18)  & \phantom{-}2.03(12)  & -0.844(29) & \phantom{-}1.29(10)  & \phantom{-}0.240(12) \\
0.9 & \phantom{-}3.05(17)  & -1.05(19)  & \phantom{-}1.90(13)  & \phantom{-}1.34(21)   & -0.26(84)   & -0.27(49)   & \phantom{-}1.52(29)  & -0.75(15)  & \phantom{-}1.24(17)  & \phantom{-}0.305(39) \\
1   & \phantom{-}3.15(59)  & -0.71(36)  & \phantom{-}1.87(30)  & \phantom{-}1.55(17)   & -0.2(1.6)   & -0.16(50)   & \phantom{-}1.98(46)  & -0.692(88) & \phantom{-}1.51(12)  & \phantom{-}0.38(19)  \\
\bottomrule\bottomrule
\end{tabular}
\caption{
    Values of the energy levels at $\beta = 8$. 
    These numbers correspond to the squares or points displayed in~\ref{fig:spectrum} and~\ref{fig:EcontPerMu} respectively.
}\label{tab:spectrum}
\end{table*}
  
\FloatBarrier

\section{Conclusions}\label{sec:conclusions}
In this work we have performed an initial Monte Carlo study of the electronic structure of a single doped perylene $\perylene$ molecule described with the Hubbard model.
We treated discretization errors by simulating at three discretizations and performing a continuum limit extrapolation. 
The effect of temperature is studied qualitatively at three values.
Central to this study is the scan over chemical potential starting at half filling $(\mu = 0)$, including the first doping transition $(0.4 < \mu < 0.5)$, and stretching further out to $\mu = 1.1$.
We quantify the doping by calculating the total system charge, providing evidence for the position of the transition.
We map out the low single particle energy spectrum at each chemical potential, backing the transition with a negative energy state moving out of the Fermi sea.
Throughout all results, we find significant divergence from the non-interacting model.
In particular, the point of transition moves to larger chemical potentials and an additional splitting of accidentally degenerate energy states emerges.
For technological applications to perylene-derived molecules we can easily leverage a more accurate interaction.
We also plan to compute charge-neutral excitations, responses to external electromagnetic sources, and to carefully study the cold regime.

\begin{acknowledgments}

We gratefully acknowledge the computing time on the supercomputer JURECA~\cite{jureca-2021} at Forschungszentrum Jülich, including the VSR grants 25188, 27702, and 30278.
This work was funded in part by the STFC Consolidated Grant ST/T000988/1, 
by the MKW NRW under the funding code NW21-024-A, 
by Deutsche Forschungsgemeinschaft (DFG, German Research Foundation) under grant “NFDI 39/1” (PUNCH4NFDI) and the CRC 1639 NuMeriQS – project no.\ 511713970,
and by RWTH Exploratory Research Space (ERS) Grant no.\ PF-JARA-SDS005.
\end{acknowledgments}

\clearpage
\appendix

\section{Analysis Details}\label{apx-sec:analysis_details}
In this appendix we describe in detail each step of the analysis.

\subsection{Reweighting}\label{subsec:Reweighting}

When dealing with systems obeying a complex valued action a way to utilize Monte Carlo integration is reweighting.
For this, the Markov Chain is generated by sampling according to the Boltzmann distribution originating from the real part of the action effectively
treating the complex phase $e^{-\im \Im{S}}$ as part of the observable. 
In order to generate the intended observables the relation
\begin{equation}
    \expval{\mathcal{O}} = \frac{ \expval{\mathcal{O} e^{-\im \Im{S}}}_{\Re{S}} }{\expval{ e^{-\im \Im{S}} }_{\Re{S}}}.
    \label{eq:reweighting}
\end{equation}
has to be evaluated.
Under a bootstrap analysis
each resample is evaluated in this way maintaining the correlations and fluctuations of the observables with the phase.

\subsection{Autocorrelation}\label{subsec:Autocorrelation}

When estimating statistical uncertainty of observables, especially with bootstrap based analysis,
the observables need to be statistically independent between configurations.
This naively is not the case for Markov Chain algorithms. 
Yet, we can ensure statistical independence by various means for example by striding -- only measuring on every $\mathrm{n}^\mathrm{th}$ trajectory with n big enough.
A post-processing option is to evaluate the autocorrelation function
\begin{equation}
\Gamma_{\O{}}(\nu) \propto \sum_{ n = 0 } ^{\Ncfg -\nu} \left( \O{\Phi_{n+\nu} } - \expval{\O{}}\right)\left( \O{\Phi_{n} } - \expval{\O{}}\right)^*
    \label{eq:autocorrelation}
\end{equation}
normalized by $\Gamma_{\O{}}(0)$,
and estimating the integrated autocorrelation time~\cite{sokal1997monte,wolff2007monte},  
\begin{equation}
    \tau_\mathrm{int}^{\O{}} = \frac{1}{2} + \sum_{\nu = 1}^{M} \Gamma_{\O{}}(\nu).
\end{equation}
One can find the cut-off $M\ll\Ncfg$ by searching for the smallest number such that $M \leq 10\cdot \tau_\mathrm{int}^{\O{}}$~\cite{sokal1997monte}.

For the analysis discussed here we measure on every $\nth{10}$ trajectory and subsequently
identify the largest autocorrelation time over all our considered observables (the set of correlators $\Csp_{x,y}(\tau)$).
To ensure no observable is autocorrelated, we use this largest integrated autocorrelation as a stride between measurements resulting in $\Ncfg^\mathrm{indep} = \nicefrac{\Ncfg}{2\max_{\O{}}\{\tau_\mathrm{int}^{\O{}}\} }$ independent samples\footnote{
    For convenience, we denote the number of independent samples simply by $\Ncfg$ from here on.
}.
We find that most of the time $\tau_\mathrm{int}^{\O{}} \approx 0.5$. In exceptional cases, we find $\tau_\mathrm{int}^{\O{}} \approx 1$.

\subsection{Spectral Decomposition}\label{subsec:effective masses}

By inserting complete sets of Hamiltonian eigenstates into the thermal trace defining the single-particle (and -hole) correlators \eqref{eq:correlator-Minv} we find the spectral decomposition
\begin{align}
    \label{eq:full spectral decomposition}
    \Csp_{xy}(\tau) 
    &=
    \frac{1}{\mathcal{Z}} \sum_{\alpha n}
        z_{\alpha x n}^{\phantom{*}} z_{\alpha y n}^*
            e^{-E_n\tau}
            e^{-E_\alpha(\beta-\tau)}\,,
    \\
    \mathcal{Z}
    &=
    \sum_{n} e^{-E_n \beta}
    \label{eq:spectral Z}
\end{align}
where we define the overlap factors
\begin{align}
z_{\alpha x n}  &=  \mel{\alpha}{p_x}{n}
    \label{eq:overlap factors}
\end{align}
and $\alpha$ and $n$ label many-body energy eigenstates that differ by the quantum numbers of a single particle.

In the large-$\beta$ limit the spectral decomposition simplifies to
\begin{align}
    \label{eq:spectral decomposition}
    \Csp_{xy}(\tau) 
    &=
    \sum_{n}
    z_{\Omega x n}^{\phantom{*}} z_{\Omega y n}^*
            e^{-(E_n-E_\Omega)\tau}
\end{align}
with $\left|\Omega\right\rangle$ the many-body ground state (if multiple states are degenerate, the decomposition is the obvious sum).
By analyzing the spectral decomposition we can find energy differences from the ground state; at finite chemical potential $\mu\neq0$ the eigenvalues $E$ are of $H-\mu Q$.

\subsection{Diagonalizing Correlators}\label{subsec:Diagonalize Correlators}

\begin{figure*}
	\includegraphics[width=\textwidth]{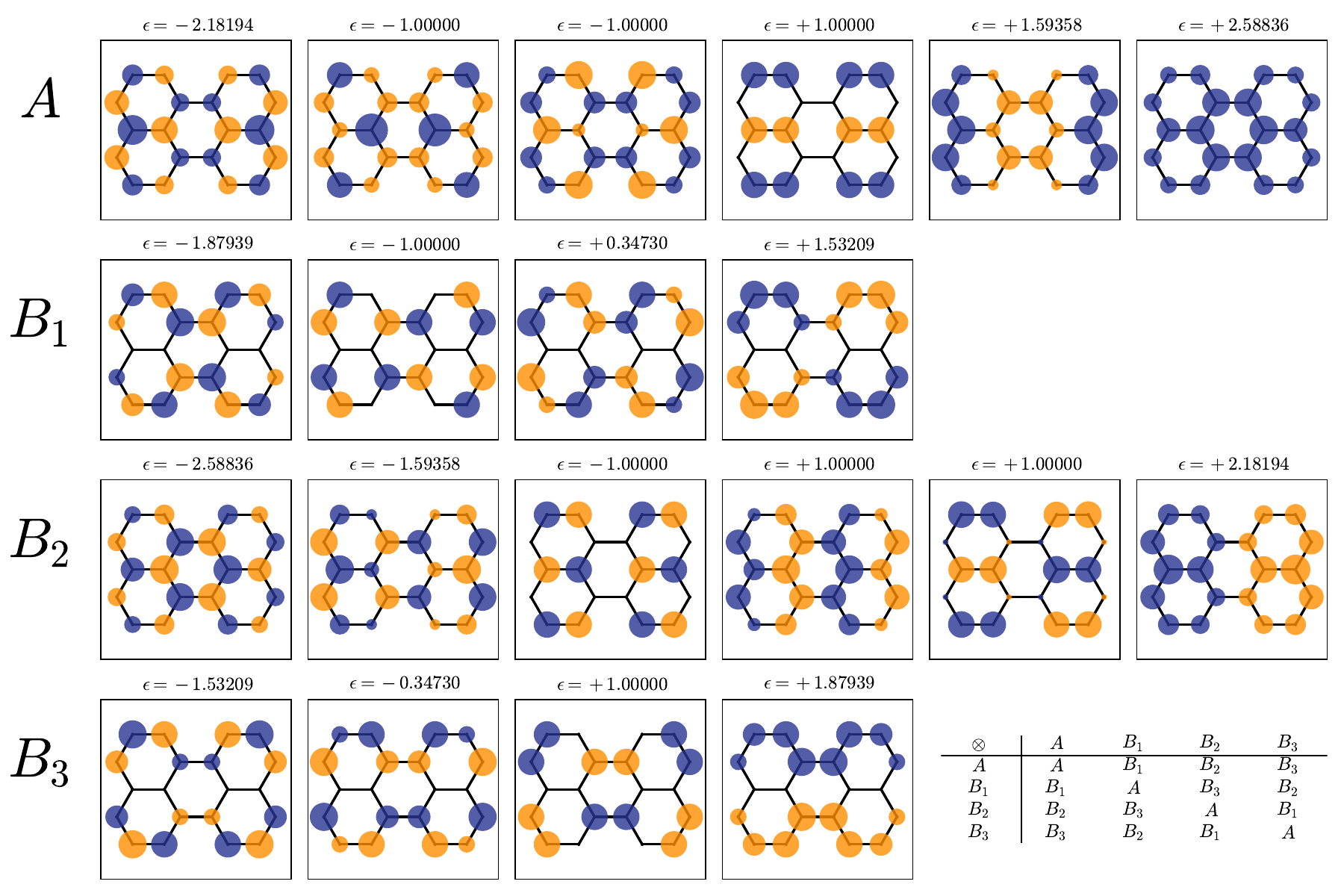}
	\caption{
		Single-particle eigenoperators $\sum_x c^*_x p^{\dagger}_x$ of the perylene tight-binding Hamiltonian.
		The circle on each site is scaled according to $|c|$, the absolute value of operator's amplitude there, and colored according to its sign (dark blue is positive, light red is negative).
		Each row is an irreducible representation, each irrep is sorted by the non-interacting tight-binding energy eigenvalue $\epsilon$ labelling the operators.
		In the lower-right we show the $D_2$ product table.
	}
	\label{fig:irreps}
\end{figure*}

An analogue of CPT symmetry allows us to average the single-particle and the time-reversed single-hole correlators; this helps us increase statistics and reduce the amount of required analysis, and we henceforth drop the single-particle superscript on $C$.

The point symmetry group of perylene is typically identified as $D_{2h}$.
Our Hamiltonian, however, treats the ions as fixed, and we can split the symmetry into the dihedral group $D_2$ and a $\mathbb{Z}_2$ whose only action is to flip spin components; we already average over particles and holes leveraging the equivalent of CPT, so this $\mathbb{Z}_2$ is accounted for.
The $D_2$ symmetry can be understood as a combination of reflections across the two principle axes; the $A$ irrep is even under both reflections, the $B_1$ irrep is odd under top-to-bottom reflections and odd under left-to-right reflections, $B_2$ is even/odd, and $B_3$ is odd/even.

We can perform a basis transformation on the correlation functions \eqref{eq:correlator-Minv} to compute correlators of $\sum_x c^*_x p^{\dagger}_x$ with the amplitudes $c$ defined on every site.
The vector space defined on the 20 sites can be decomposed into invariant subspaces on which the action of the $D_2$ symmetries act irreducibly as $A$, $B_1$, $B_2$, and $B_3$; in a slight but common abuse of language we identify these invariant subspaces as the irreps themselves.
The irreps are all one-dimensional and have multiplicity 6, 4, 6, and 4, respectively.
Fig.~\ref{fig:irreps} shows an orthonormal basis of operators for each irrep, chosen to diagonalize the tight-binding ($U=0$) problem.

We can divide the ions of the lattice in A and B sublattices such that neighbours are always in the different sublattice.
If we multiply all the fermion operators on a single sublattice by $-1$, the tight-binding Hamiltonian flips, because every possible hopping picks up exactly one sign, and we see that the tight-binding spectrum is symmetric around zero.

However, this sublattice symmetry does not commute with the $D_2$ point group, so the operators with opposite tight-binding energies (related by staggering the amplitudes' signs on one sublattice) appear in different irreps; this is particularly clear in Fig.~\ref{fig:irreps} for the $B_1$ and $B_3$ irreps which have no accidental degeneracies.
Another good example is the highest-energy $A$ operator (with uniformly-signed amplitudes) and the lowest-energy $B_2$ operator (with corresponding staggered amplitudes).
The ion-independent Hubbard interaction does not break the $D_2$ symmetry.

We can use the amplitudes $c$ to construct a unitary matrix that block-diagonalizes the correlator $C$,
\begin{align}
    C_{\Lambda_i',\Lambda^{}_j}(\tau) &= \sum_{xy} U_{\Lambda'_i,x} C_{xy}(\tau) (U^\dagger)_{y,\Lambda^{}_j} = C^\Lambda_{ij}(\tau) \delta_{\Lambda'\Lambda}
	\label{eq:block diagonalize}
\end{align}
where $\Lambda$ and $\Lambda'$ label the $D_2$ irreps and $i$ and $j$ operators of the respective irrep.
Because our Hamiltonian has $D_2$ symmetry the irrep is conserved and the transformed correlator is block diagonal, as shown in the second equality \eqref{eq:block diagonalize}.
Each block $C^\Lambda(\tau)$ has a spectral decomposition \eqref{eq:spectral decomposition} which sums over only states $n$ that differ from the ground state by irrep $\Lambda$; put another way in the full spectral decomposition \eqref{eq:full spectral decomposition} the $D_2$ Wigner-Eckhart theorem states that $\alpha = \Lambda \otimes n$ using the $D_2$ product table in Fig.~\ref{fig:irreps} where $\alpha$ and $n$ are the irreps of their respective states.

When the interaction is weak, the single-particle correlation function transformed into this basis is nearly diagonal because the basis diagonalizes the tight-binding problem; when the interaction is strong, it remains block diagonal in irrep but within an irrep the operators can mix.
Because every off-diagonal entry has differing contributions from excited states, no single unitary transformation diagonalizes an irreducible block for every time $\tau$.
We can nevertheless diagonalize each time slice independently.

Many diagonalization routines sort eigenvalues, which can lead to misidentifying the time dependence when correlators cross and cause trouble under a bootstrap analysis.
A variety of sorting  methods that can help to avoid this misidentification are discussed in Ref.~\cite{fischer2020generalised}.
To maintain the ordering of states and avoid said ambiguity, we diagonalize using a Jacobi method based on Givens rotation: the largest off-diagonal elements are iteratively rotated into the diagonal.
By tracking these rotations we can also find the linear combination of operators that yield a diagonalized time slice.

However, this tracking procedures fail when correlators within an irrep cross; if we diagonalize timeslice-by-timeslice the crossings have level repulsion and introduce unphysical discontinuities in the resulting correlators.
These crossings frequently appear, rendering a perfect timeslice-by-timeslice diagonalization inaccessible.
This numerical problem stems from using only the 20 single particle operators, which do not constitute a complete basis of the spin-half $Q=1$ sector.
For example, we do not include in our calculation operators which have the same quantum numbers as our single-particle interpolators, like $p^\dagger h^\dagger h$.
Interacting eigenstates mix $p^\dagger$ with all such operators, but our irreducible blocks are truncated to only the single-particle interpolators.
If we would measure a much bigger correlator built from a complete basis of the single-particle sector the timeslice-by-timeslice diagonalization would produce perfect correlators with no repulsion.

Rather than grapple with these discontinuities, we instead adopt a variational approach.  
Given $N_t$ unitaries $U_t$, one for each timeslice, we select the one that best diagonalizes all other time slices,
\begin{equation}
\tau = \min_{t\neq t'} \norm{ U_{t}^\dagger \cdot U_{t'}  - \mathds{1} },
\end{equation}
and use it to approximately diagonalize the blocks.
This unitary can be thought of as variationally selecting a linear combination of the tight-binding eigenoperators shown in figure~\ref{fig:irreps}.
From these mostly-diagonalized blocks we simply take the diagonal elements,
resulting in a set of 20 correlators $C_{\Lambda_i}(\tau)$ where $\Lambda$ labels an irrep and $i$ is just an index.
From these variationally-diagonalized correlators we are ultimately interested in the lowest energy---or more precisely, the energy closest to zero---in the spectral decomposition \eqref{eq:spectral decomposition}.

\subsection{Fitting Energies}\label{subsec:Fitting}

In order to systematically reduce the effect of excited states, we can fit correlators to a truncated spectral decomposition.
The fit program proceeds with three steps; 
First, decide on a fit model, including number of states -- terms in~\eqref{eq:spectral decomposition} -- and fit range as well as identify prior-knowledge;
second perform a Bayesian fit;
and last measure how well the fit did.

As mentioned before the spectrum contains positive and negative energies.
Therefore, the spectral decomposition can be split into two contributions, decaying ($z_n^L,E_n^L$) and increasing exponentials ($z_n^R, E_n^R$), suppressing the state label $\Lambda_i$ for clarity.
To further stabilize the fit and ensuring that $E_0^{L/R}$ is the smallest energy, the model is recast with relative energy differences 
$E_n^{R/L} \to \Delta E_n^{R/L}$ such that $\Delta E_n^{L/R} = E_n^{L/R} - E_{n-1}^{L/R} > 0$ resulting in the fit model~\eqref{eq:exp-model}
With this fit model, and the variational basis constructed in the previous section, we can identify the energy gap and overlap by
\begin{align}
    E^{\Lambda_i}_{\phantom{0}} = E_0^{L} \text{ or } -E_0^{R}, \\
    \abs{z_{\Omega \Lambda_i}}^2 = z_0^{L} \text{ or } z_0^{R}.
\end{align}
If the correlator $C_{\Lambda_i}(\tau)$ is primarily decaying take $z_0^L,E_0^L$ otherwise $z_0^R,E_0^R$. 
This choice is made based on the fact that the slowest decay/increase of the correlator comes from the lowest energy, consequently we treat the other as excited state contamination.
We truncate the spectral decomposition \eqref{eq:exp-model} after $\Nexp = 1,2$ on the longer part of the correlator and keep $\Nexp=1$ on the shorter end.

The contribution from excited states is different from time slice to time slice. 
Thus, it is advisable to include different fit intervals $\tau \in \delta [\tau_\mathrm{start}, \tau_\mathrm{end}]$.
These are chosen by identifying the minimal point of the correlator, $\tau_\mathrm{min} = \min\limits_{\tau}(\abs{C_{\Lambda_i}(\tau)})$ and taking all possible combinations of 
$\tau_\mathrm{start} < \tau_\mathrm{min} < \tau_\mathrm{end}$.
For many correlators, the center part is relatively flat due to overlaps of exponentials causing artificially small energies $E_0^{L/R}$.
To prevent this behaviour, the space of fit intervals is truncated to always take at least 75\% of the subintervals to the left and right, i.e.
$\tau_\mathrm{start} < 0.75 \cdot ( \tau_\mathrm{min} - 1 )$, 
$0.75 \cdot ( \Nt-1 - \tau_\mathrm{min} ) < \tau_\mathrm{end}$. 

The last ingredients are the priors to the fit. 
As discussed previously, the non-interacting energy spectrum can be accessed analytically through $\epsilon_{\Lambda_i}$.
Though we expect divergence from this, it at least provides a good order of magnitude of the energies of the interacting simulations.
Therefore, we use this information in combination with a log-normal prior for the 1-state fits,
\begin{align}
    E_0^{L/R} &\sim \log\mathcal{N}\left(\abs{E^{\Lambda_i}_{U=0}(\mu)} , \ \abs{E^{\Lambda_i}_{U=0}(\mu)} \right)
\end{align}
In case a zero crossing is expected $(E_{U=0}^{\Lambda_i}(\mu) = 0)$, we simply use a gaussian prior with mean 0 and standard deviation 10.
Considering the form of the correlator, especially its magnitudes at the end, we expect that the overlaps are $\order{1}$. 
This is encoded with a gaussian-prior with mean and standard deviation equal 1.
The variationally-diagonalized correlators are positive-definite, so too large a standard deviation would allow unphysical results.

For two-state fits the priors are partially determined by the one-state fit results we have already obtained.
We utilize the model average, discussed in the next section~\ref{subsec:Model Averaging}. 
The central value serves as a mean to the (log-)gaussian prior while the standard deviation is determined by the maximum of $5\sigma$ and $10\%$ of the central value, giving the fitter enough freedom to adjust the fit result.
For the first two-state fit two additional parameter $z_1^{L/R}, \Delta E_1^{L/R}$, that can not be obtained from the one state fit results, are using a flat prior.

This fitting procedure is done on the central values of the correlator to provide central values for the energies.
Furthermore, it is performed on each bootstrap sample to provide uncertainties on the energies.
The fits are done in an uncorrelated manner, as the correlation is being tracked through the bootstraps.

\subsection{Model Averaging}\label{subsec:Model Averaging}

This procedure results in a high number of fits obtained using \texttt{lsqfit}~\cite{lsqfit}. 
For each, we compute the Akaike information criterion~\cite{jay2021bayesian,neil2024improved,neil2023model} 
\begin{equation}\label{eq:akaike}
    \AIC = \chi^2 + 2 \Nparams - 2 \left\vert\tau_e - \tau_s\right\vert,
\end{equation}
This measure penalizes the number of parameters and smaller fit range which is exactly what we are varying.
A thorough discussion on this criterion in comparison to others can be found in Ref.~\cite{neil2024improved}.
With that we weight each fit result by the associated probability
\begin{equation}
    P\left( \mathrm{model} \vert \mathrm{data} \right) \propto e^{-\frac{1}{2} \AIC}.
\end{equation}
to obtain the final parameter value $\expval{p^{\phantom{i}}_n}$, $p^i_n\in\{z_n^{L/R}, E_n^{L/R}\}_{n=0}^{\Nexp-1}$ where $i$ labels the different results,
\begin{equation}
    \expval{p_n} = \frac{
        \sum_{i} e^{-\frac{1}{2} \AIC_i} p^i_n
    }{
        \sum_{i} e^{-\frac{1}{2} \AIC_i}
    }.\label{eq:model-average}
\end{equation}

\subsection{Continuum Limit}\label{subsec:Limits}

\begin{figure*}
    \resizebox{!}{0.9\textheight}{\includegraphics{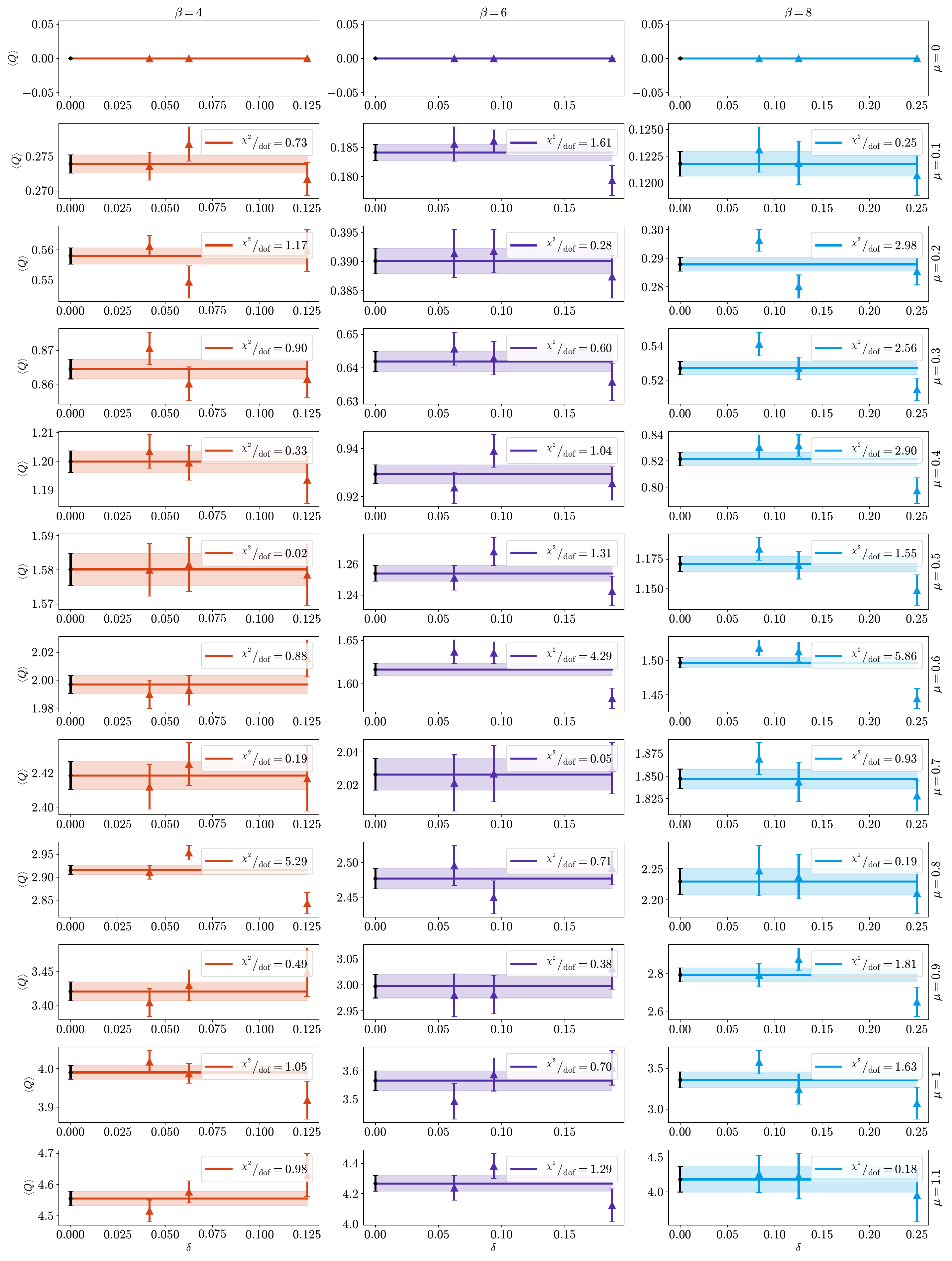}}
    \caption{ 
        Continuum limit for the total system charge $Q$.
    }\label{fig:continuumCharge}
\end{figure*}

Once the charges and model averaged energies for a given set of parameters $(\Nt,\beta,\mu)$ are obtained a continuum limit has to be performed, $\delta=\nicefrac{\beta}{\Nt} \to 0$.
The temperatures considered are too high for a reliable zero-temperature limit.
We follow a similar approach as outlined in~\cite{ostmeyer2020semimetal}. 
Expanding the correlator~\eqref{eq:correlator-Minv} in a geometric sum and expanding in small $\delta$ suggests a polynomial in $\delta$. 
This results in a expansion for the total charge, estimated from the $\nth{0}$ time slice,
\begin{equation}
    \expval{Q(\delta,\beta)} = Q_0(\beta) + \sum_{d=1}^{D} \delta^d Q_d(\beta) + \order{\delta^D}
\end{equation}
Usually, a control point is beneficial as otherwise priors can strongly bias fits of this form leaving us with $D=1$ (2 parameters).
Following the string of chemical potentials, the slopes $Q_1$ are distributed without a clear trend suggesting that discretization effects can be neglected -- we are deep into the scaling regime.
Consequently, we perform the continuum limit with only the constant piece, $D=0$.
figure~\ref{fig:continuumCharge} provides an overview of the continuum limits for the total system charges discussed in section~\ref{sub-sec:charge}.
We find good fits across all systems, with some divergence on the coarsest lattices ($\Nt = 32$).

Developing this expansion into the spectral expansion of the correlator maintains this relation.
\begin{equation}
    E_0^{\Lambda_i}(\delta,\beta) = E_0^{\Lambda_i}(\beta) + \sum_{d=1}^{D} \delta^d E_d^{\Lambda_i}(\beta) + \order{\delta^D}
\end{equation}
Where this sum is truncated to some power D.
We truncate at $D=0$ similar to the total charge.

In figure~\ref{fig:continuumEnergy} the continuum limit for $B_3^3$ is shown. 
This flat extrapolation is typical extrapolation for all states and we do not show them here.
The results are further summarized in table~\ref{tab:spectrum_beta4} for $\beta = 4$ and~\ref{tab:spectrum_beta6} for $\beta = 6$.

\section{More Spectrum}\label{apx-sec:moreSpectrum}
\begin{figure*}
    \centering
    \resizebox{!}{0.9\textheight}{\includegraphics{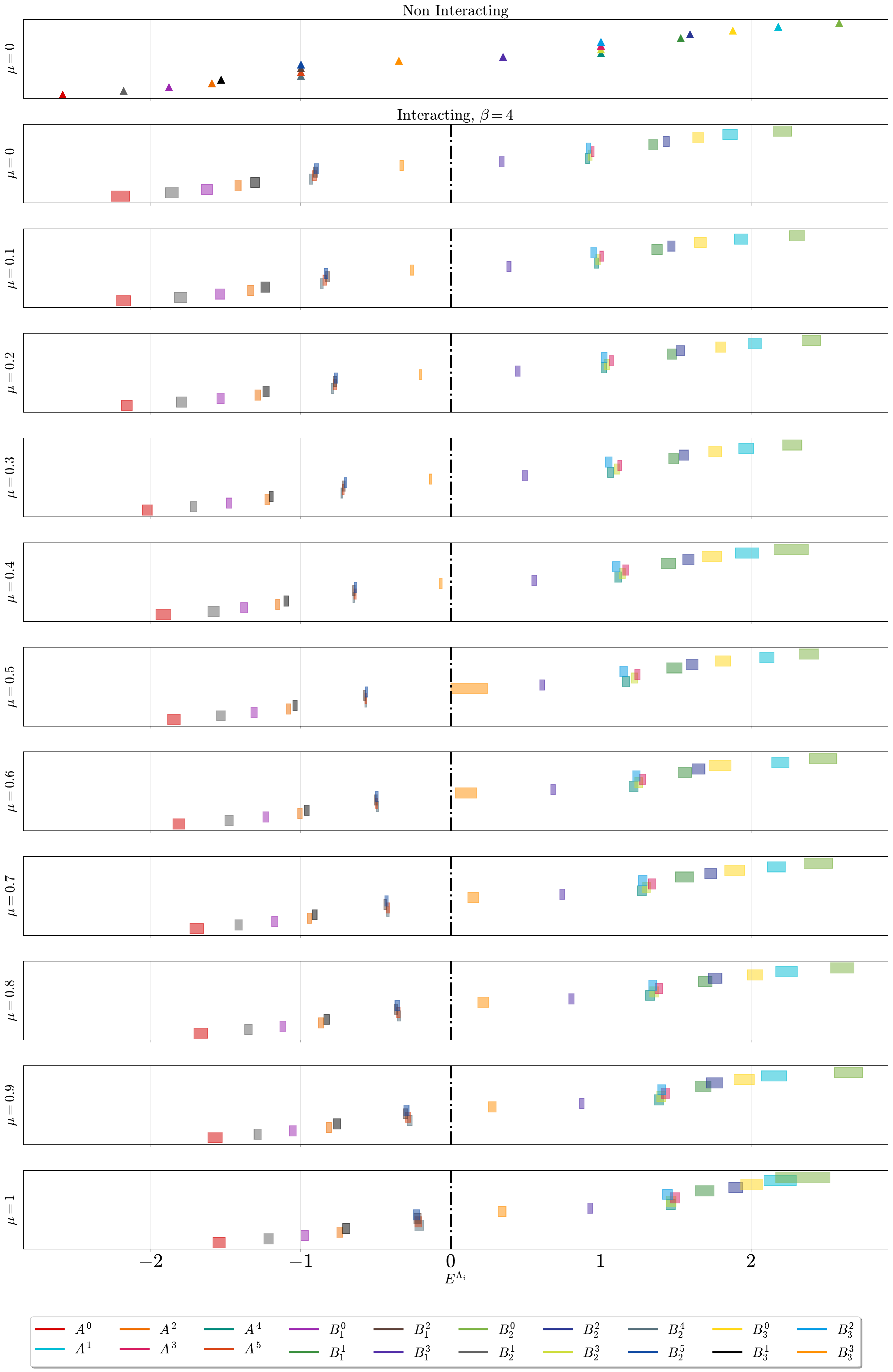}}
    \caption{
        Similar spectrum as in figure~\ref{fig:spectrum} with $\beta = 4$.
        The sign problem is significantly less sever than at $\beta = 8$ consequently giving better estimates past $\mu = 0.8$.
    }\label{fig:spectrum_beta4}
\end{figure*}

\begin{figure*}
    \centering
    \resizebox{!}{0.9\textheight}{\includegraphics{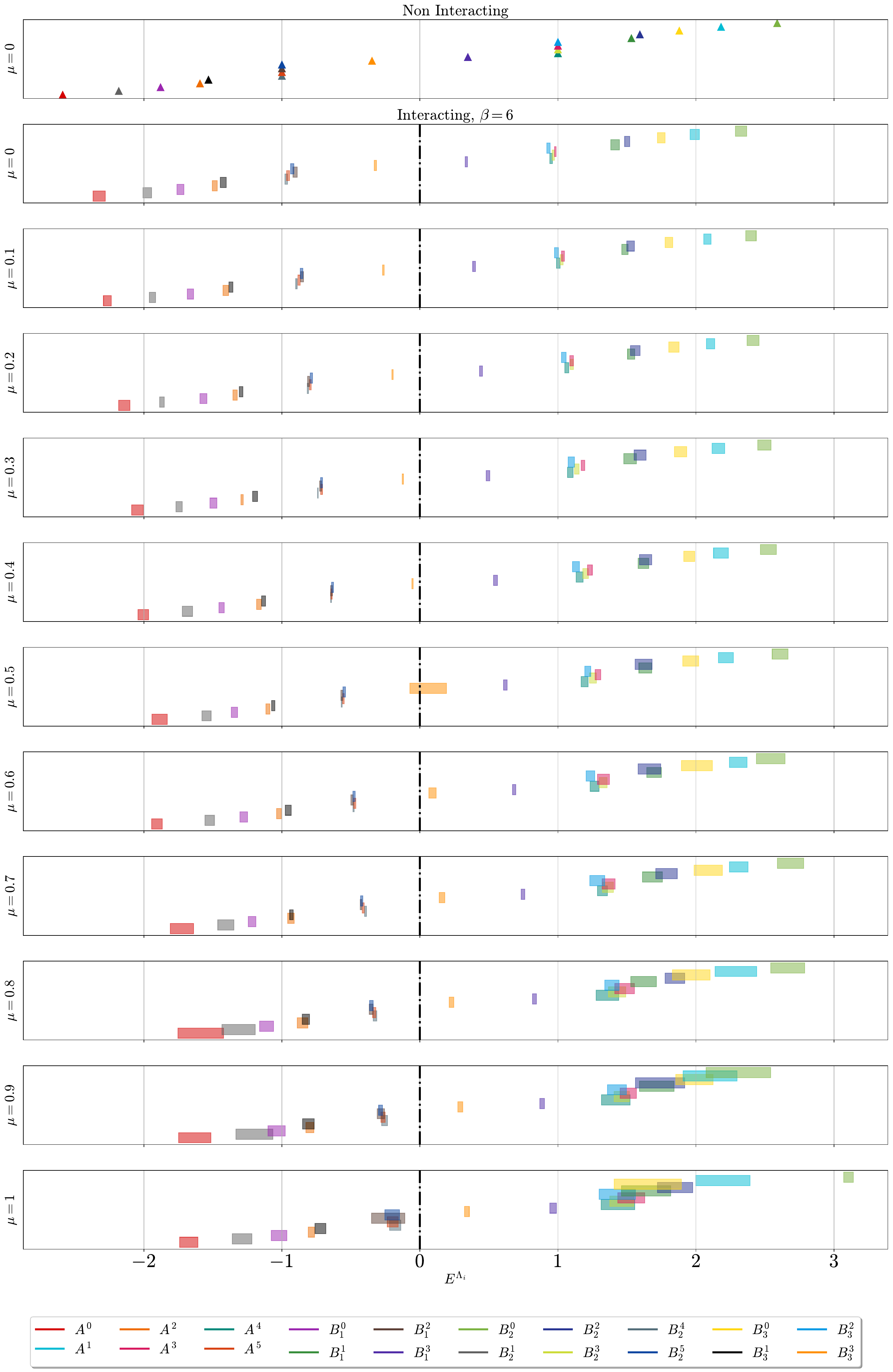}}
    \caption{
        Similar spectrum as in figure~\ref{fig:spectrum} with different $\beta = 6$.
        The sign problem is less severe than at $\beta = 8$ consequently giving better estimates past $\mu = 0.8$.
    }\label{fig:spectrum_beta6}
\end{figure*}

We provide the $\beta = 4,6$ spectra in figures~\ref{fig:spectrum_beta4} and~\ref{fig:spectrum_beta6} and 
summarize the values in the tables~\ref{tab:spectrum_beta4} and~\ref{tab:spectrum_beta6}
respectively.

\begin{figure*}
    \centering
    \resizebox{\textwidth}{!}{\includegraphics{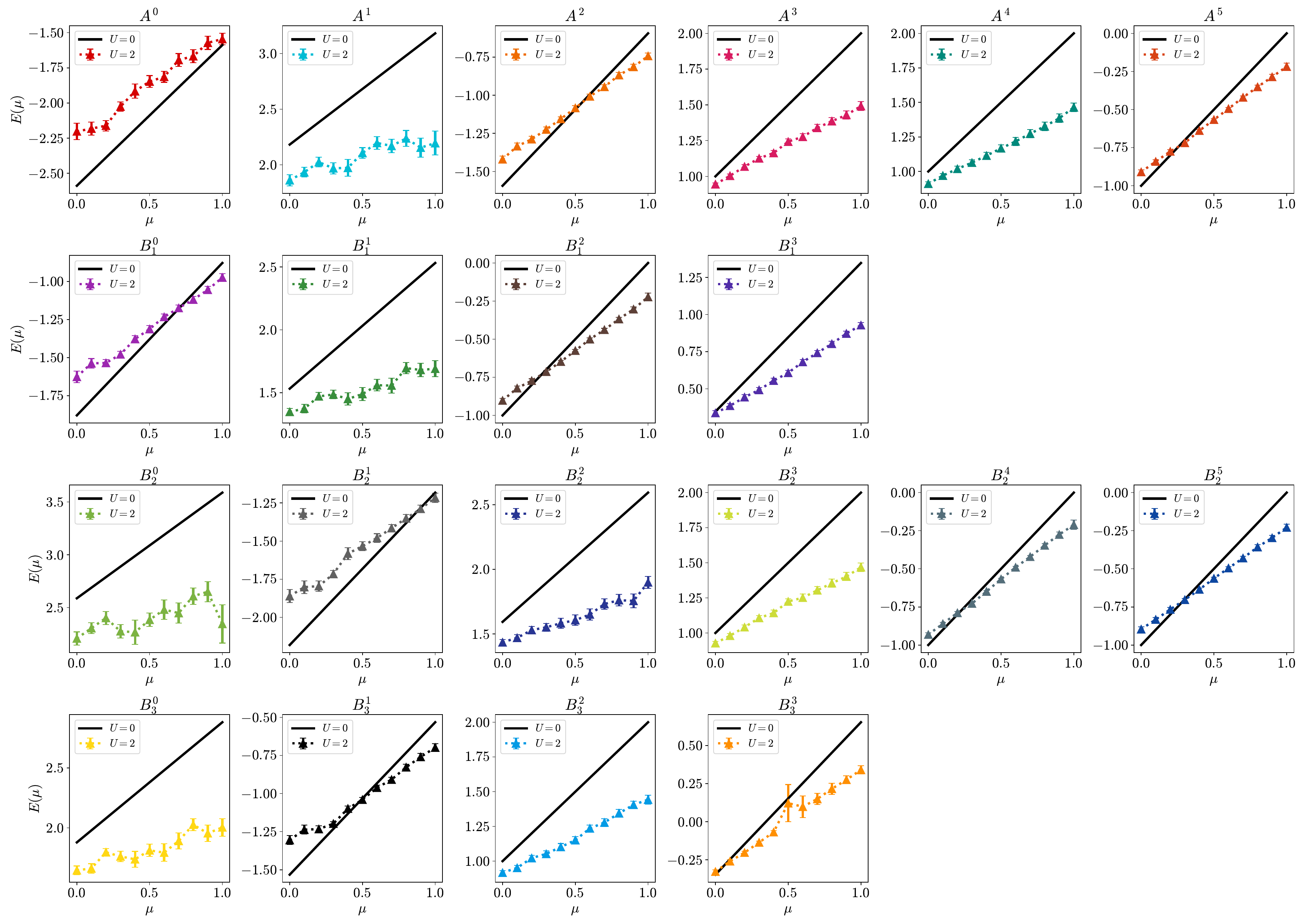}}
    \caption{
        Similar plots as in figure~\ref{fig:EcontPerMu} with different $\beta = 4$.
    }\label{fig:EcontPerMu_beta4}
\end{figure*}

\begin{table*}
\begin{tabular}{l|llllllllll}
\toprule\toprule
    \multicolumn{1}{c}{$\mu$}   &  
    \multicolumn{1}{c}{$A^0$}   & 
    \multicolumn{1}{c}{$A^1$}   & 
    \multicolumn{1}{c}{$A^2$}   & 
    \multicolumn{1}{c}{$A^3$}   & 
    \multicolumn{1}{c}{$A^4$}   & 
    \multicolumn{1}{c}{$A^5$}   & 
    \multicolumn{1}{c}{$B_1^0$} & 
    \multicolumn{1}{c}{$B_1^1$} & 
    \multicolumn{1}{c}{$B_1^2$} & 
    \multicolumn{1}{c}{$B_1^3$} \\[1em]
0   & -2.202(59) & \phantom{-}1.861(49) & -1.420(21) & \phantom{-}0.945(10) & \phantom{-}0.911(15) & -0.910(15)  & -1.627(37) & \phantom{-}1.348(28) & -0.902(15)  & \phantom{-}0.338(16) \\
0.1 & -2.182(46) & \phantom{-}1.933(44) & -1.334(21) & \phantom{-}1.004(12) & \phantom{-}0.970(16) & -0.841(12)  & -1.537(31) & \phantom{-}1.373(35) & -0.822(14)  & \phantom{-}0.386(14) \\
0.2 & -2.161(37) & \phantom{-}2.025(44) & -1.288(19) & \phantom{-}1.068(14) & \phantom{-}1.019(18) & -0.775(11)  & -1.535(24) & \phantom{-}1.471(31) & -0.772(13)  & \phantom{-}0.444(16) \\
0.3 & -2.025(33) & \phantom{-}1.968(49) & -1.224(16) & \phantom{-}1.126(14) & \phantom{-}1.064(21) & -0.7187(61) & -1.479(19) & \phantom{-}1.486(34) & -0.7135(89) & \phantom{-}0.493(17) \\
0.4 & -1.916(50) & \phantom{-}1.973(76) & -1.155(15) & \phantom{-}1.163(19) & \phantom{-}1.115(23) & -0.6385(77) & -1.378(23) & \phantom{-}1.450(49) & -0.6473(79) & \phantom{-}0.556(16) \\
0.5 & -1.847(42) & \phantom{-}2.106(48) & -1.083(14) & \phantom{-}1.243(18) & \phantom{-}1.168(25) & -0.5663(66) & -1.313(22) & \phantom{-}1.489(51) & -0.5750(78) & \phantom{-}0.608(16) \\
0.6 & -1.814(39) & \phantom{-}2.197(58) & -1.007(15) & \phantom{-}1.277(22) & \phantom{-}1.218(30) & -0.4947(74) & -1.233(20) & \phantom{-}1.559(46) & -0.5001(91) & \phantom{-}0.681(16) \\
0.7 & -1.694(46) & \phantom{-}2.170(59) & -0.944(14) & \phantom{-}1.339(24) & \phantom{-}1.274(30) & -0.420(10)  & -1.175(20) & \phantom{-}1.557(60) & -0.4380(91) & \phantom{-}0.742(16) \\
0.8 & -1.667(46) & \phantom{-}2.237(71) & -0.868(18) & \phantom{-}1.386(26) & \phantom{-}1.327(31) & -0.351(13)  & -1.120(20) & \phantom{-}1.695(45) & -0.368(11)  & \phantom{-}0.803(18) \\
0.9 & -1.572(48) & \phantom{-}2.154(85) & -0.813(18) & \phantom{-}1.429(28) & \phantom{-}1.385(31) & -0.286(16)  & -1.055(23) & \phantom{-}1.681(53) & -0.303(16)  & \phantom{-}0.873(16) \\
1   & -1.546(41) & \phantom{-}2.19(11)  & -0.742(18) & \phantom{-}1.491(31) & \phantom{-}1.464(31) & -0.218(23)  & -0.975(25) & \phantom{-}1.691(64) & -0.223(24)  & \phantom{-}0.929(17) \\
\midrule
                                &
    \multicolumn{1}{c}{$B_2^0$} & 
    \multicolumn{1}{c}{$B_2^1$} & 
    \multicolumn{1}{c}{$B_2^2$} & 
    \multicolumn{1}{c}{$B_2^3$} & 
    \multicolumn{1}{c}{$B_2^4$} & 
    \multicolumn{1}{c}{$B_2^5$} & 
    \multicolumn{1}{c}{$B_3^0$} & 
    \multicolumn{1}{c}{$B_3^1$} & 
    \multicolumn{1}{c}{$B_3^2$} & 
    \multicolumn{1}{c}{$B_3^3$} \\[1em]
0   & \phantom{-}2.208(62) & -1.861(43) & \phantom{-}1.435(21) & \phantom{-}0.927(13) & -0.932(12)  & -0.896(15)  & \phantom{-}1.647(36) & -1.305(30) & \phantom{-}0.917(14) &           -0.328(12)  \\
0.1 & \phantom{-}2.306(51) & -1.802(42) & \phantom{-}1.470(25) & \phantom{-}0.980(15) & -0.8605(94) & -0.833(13)  & \phantom{-}1.663(39) & -1.237(29) & \phantom{-}0.952(19) &           -0.2596(96) \\
0.2 & \phantom{-}2.402(62) & -1.796(35) & \phantom{-}1.530(29) & \phantom{-}1.041(17) & -0.7890(90) & -0.766(12)  & \phantom{-}1.797(31) & -1.233(21) & \phantom{-}1.022(19) &           -0.2026(87) \\
0.3 & \phantom{-}2.276(63) & -1.715(23) & \phantom{-}1.552(31) & \phantom{-}1.107(16) & -0.7275(52) & -0.7031(93) & \phantom{-}1.762(44) & -1.198(13) & \phantom{-}1.052(22) &           -0.1359(84) \\
0.4 & \phantom{-}2.27(11)  & -1.582(39) & \phantom{-}1.583(38) & \phantom{-}1.142(21) & -0.6488(54) & -0.6356(87) & \phantom{-}1.739(66) & -1.099(16) & \phantom{-}1.102(25) &           -0.068(11)  \\
0.5 & \phantom{-}2.385(64) & -1.533(29) & \phantom{-}1.607(40) & \phantom{-}1.224(20) & -0.5674(62) & -0.5630(88) & \phantom{-}1.813(52) & -1.040(14) & \phantom{-}1.151(25) & \phantom{-}0.12(12)   \\
0.6 & \phantom{-}2.481(91) & -1.480(28) & \phantom{-}1.650(43) & \phantom{-}1.251(26) & -0.4904(73) & -0.4955(96) & \phantom{-}1.794(73) & -0.962(16) & \phantom{-}1.236(25) & \phantom{-}0.098(71)  \\
0.7 & \phantom{-}2.449(95) & -1.415(24) & \phantom{-}1.732(41) & \phantom{-}1.304(26) & -0.4208(92) & -0.430(11)  & \phantom{-}1.892(66) & -0.908(16) & \phantom{-}1.278(29) & \phantom{-}0.149(37)  \\
0.8 & \phantom{-}2.608(78) & -1.350(25) & \phantom{-}1.761(45) & \phantom{-}1.355(29) & -0.346(13)  & -0.357(16)  & \phantom{-}2.027(50) & -0.828(19) & \phantom{-}1.345(27) & \phantom{-}0.216(37)  \\
0.9 & \phantom{-}2.651(95) & -1.288(24) & \phantom{-}1.756(54) & \phantom{-}1.402(29) & -0.275(17)  & -0.297(17)  & \phantom{-}1.955(68) & -0.760(24) & \phantom{-}1.406(26) & \phantom{-}0.275(26)  \\
1   & \phantom{-}2.35(18)  & -1.216(31) & \phantom{-}1.899(47) & \phantom{-}1.468(32) & -0.211(30)  & -0.229(21)  & \phantom{-}2.004(73) & -0.698(24) & \phantom{-}1.443(33) & \phantom{-}0.340(26)  \\
\bottomrule\bottomrule
\end{tabular}
\caption{
    Values of the energy levels at $\beta = 4$. 
    These numbers correspond to the squares or points displayed in~\ref{fig:spectrum_beta4} and~\ref{fig:EcontPerMu_beta4} respectively.
}\label{tab:spectrum_beta4}
\end{table*}

\begin{figure*}
    \centering
    \resizebox{\textwidth}{!}{\includegraphics{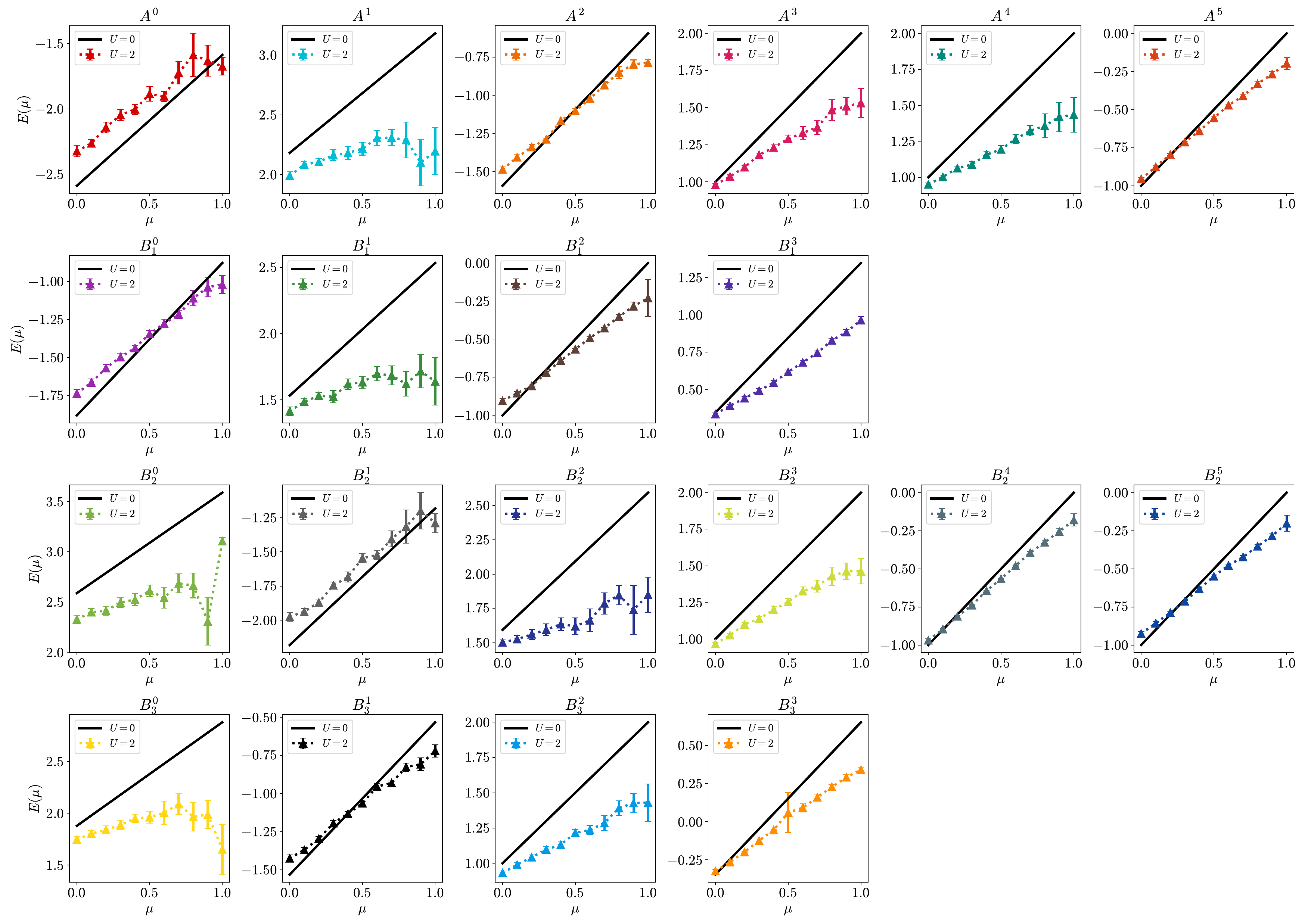}}
    \caption{
        Similar plots as in figures~\ref{fig:EcontPerMu},~\ref{fig:spectrum_beta4} with different $\beta = 6$.
    }\label{fig:EcontPerMu_beta6}
\end{figure*}

\begin{table*}
\begin{tabular}{l|llllllllll}
\toprule\toprule
    \multicolumn{1}{c}{$\mu$}   &  
    \multicolumn{1}{c}{$A^0$}   & 
    \multicolumn{1}{c}{$A^1$}   & 
    \multicolumn{1}{c}{$A^2$}   & 
    \multicolumn{1}{c}{$A^3$}   & 
    \multicolumn{1}{c}{$A^4$}   & 
    \multicolumn{1}{c}{$A^5$}   & 
    \multicolumn{1}{c}{$B_1^0$} & 
    \multicolumn{1}{c}{$B_1^1$} & 
    \multicolumn{1}{c}{$B_1^2$} & 
    \multicolumn{1}{c}{$B_1^3$} \\[1em]
0   & -2.324(44) & \phantom{-}1.991(34) & -1.486(17)  & \phantom{-}0.9809(67) & \phantom{-}0.9507(93) & -0.9555(98) & -1.735(26) & \phantom{-}1.414(31) & -0.904(15)  & \phantom{-}0.3368(88)\\
0.1 & -2.264(29) & \phantom{-}2.083(27) & -1.407(20)  & \phantom{-}1.036(11)  & \phantom{-}1.003(13)  & -0.8759(77) & -1.663(22) & \phantom{-}1.486(23) & -0.855(11)  & \phantom{-}0.393(10) \\
0.2 & -2.142(41) & \phantom{-}2.106(30) & -1.339(16)  & \phantom{-}1.099(13)  & \phantom{-}1.064(14)  & -0.7941(69) & -1.569(25) & \phantom{-}1.531(26) & -0.8071(64) & \phantom{-}0.443(12) \\
0.3 & -2.047(42) & \phantom{-}2.163(45) & -1.2889(87) & \phantom{-}1.182(13)  & \phantom{-}1.089(20)  & -0.7137(70) & -1.497(26) & \phantom{-}1.523(46) & -0.7194(76) & \phantom{-}0.493(14) \\
0.4 & -2.005(39) & \phantom{-}2.181(54) & -1.167(17)  & \phantom{-}1.232(17)  & \phantom{-}1.157(25)  & -0.6405(44) & -1.437(19) & \phantom{-}1.619(39) & -0.6403(77) & \phantom{-}0.547(14) \\
0.5 & -1.886(55) & \phantom{-}2.217(54) & -1.101(15)  & \phantom{-}1.290(20)  & \phantom{-}1.194(26)  & -0.5551(62) & -1.344(23) & \phantom{-}1.633(46) & -0.5662(73) & \phantom{-}0.618(13) \\
0.6 & -1.905(38) & \phantom{-}2.306(63) & -1.021(17)  & \phantom{-}1.330(43)  & \phantom{-}1.266(33)  & -0.4720(78) & -1.277(28) & \phantom{-}1.696(53) & -0.4917(82) & \phantom{-}0.683(13) \\
0.7 & -1.724(85) & \phantom{-}2.309(68) & -0.934(24)  & \phantom{-}1.367(47)  & \phantom{-}1.323(36)  & -0.4106(67) & -1.217(27) & \phantom{-}1.684(72) & -0.4274(58) & \phantom{-}0.746(13) \\
0.8 & -1.59(17)  & \phantom{-}2.29(15)  & -0.850(38)  & \phantom{-}1.483(72)  & \phantom{-}1.358(82)  & -0.330(12)  & -1.111(50) & \phantom{-}1.620(92) & -0.352(15)  & \phantom{-}0.829(14) \\
0.9 & -1.63(12)  & \phantom{-}2.10(20)  & -0.797(29)  & \phantom{-}1.509(59)  & \phantom{-}1.42(10)   & -0.268(15)  & -1.038(62) & \phantom{-}1.72(13)  & -0.284(27)  & \phantom{-}0.885(17) \\
1   & -1.675(67) & \phantom{-}2.20(20)  & -0.787(22)  & \phantom{-}1.531(97)  & \phantom{-}1.43(12)   & -0.197(39)  & -1.021(58) & \phantom{-}1.64(18)  & -0.23(12)   & \phantom{-}0.965(24) \\
\midrule
                                &
    \multicolumn{1}{c}{$B_2^0$} & 
    \multicolumn{1}{c}{$B_2^1$} & 
    \multicolumn{1}{c}{$B_2^2$} & 
    \multicolumn{1}{c}{$B_2^3$} & 
    \multicolumn{1}{c}{$B_2^4$} & 
    \multicolumn{1}{c}{$B_2^5$} & 
    \multicolumn{1}{c}{$B_3^0$} & 
    \multicolumn{1}{c}{$B_3^1$} & 
    \multicolumn{1}{c}{$B_3^2$} & 
    \multicolumn{1}{c}{$B_3^3$} \\[1em]
0   & \phantom{-}2.327(41) & -1.976(32) & \phantom{-}1.502(19) & \phantom{-}0.9671(82) & -0.9688(89) & -0.925(12)  & \phantom{-}1.749(29) & -1.425(21) & \phantom{-}0.931(13) &           -0.3233(91) \\
0.1 & \phantom{-}2.399(38) & -1.938(24) & \phantom{-}1.526(27) & \phantom{-}1.027(11)  & -0.8959(57) & -0.857(10)  & \phantom{-}1.804(28) & -1.369(15) & \phantom{-}0.989(15) &           -0.2649(62) \\
0.2 & \phantom{-}2.414(44) & -1.869(17) & \phantom{-}1.560(36) & \phantom{-}1.100(12)  & -0.8120(54) & -0.7867(81) & \phantom{-}1.841(37) & -1.296(13) & \phantom{-}1.042(17) &           -0.1988(44) \\
0.3 & \phantom{-}2.496(49) & -1.745(24) & \phantom{-}1.595(43) & \phantom{-}1.137(17)  & -0.7390(30) & -0.7130(70) & \phantom{-}1.889(45) & -1.194(18) & \phantom{-}1.097(22) &           -0.1245(48) \\
0.4 & \phantom{-}2.524(58) & -1.684(37) & \phantom{-}1.635(45) & \phantom{-}1.200(20)  & -0.6436(36) & -0.6326(70) & \phantom{-}1.952(40) & -1.132(15) & \phantom{-}1.131(26) &           -0.0535(56) \\
0.5 & \phantom{-}2.610(58) & -1.546(33) & \phantom{-}1.620(62) & \phantom{-}1.253(25)  & -0.5659(38) & -0.5485(85) & \phantom{-}1.962(57) & -1.063(13) & \phantom{-}1.216(21) & \phantom{-}0.06(13)   \\
0.6 & \phantom{-}2.54(10)  & -1.523(34) & \phantom{-}1.663(83) & \phantom{-}1.327(28)  & -0.4799(47) & -0.4776(88) & \phantom{-}2.01(11)  & -0.954(22) & \phantom{-}1.236(32) & \phantom{-}0.092(27)  \\
0.7 & \phantom{-}2.685(95) & -1.406(59) & \phantom{-}1.786(78) & \phantom{-}1.362(40)  & -0.3944(81) & -0.4224(85) & \phantom{-}2.09(10)  & -0.930(13) & \phantom{-}1.285(55) & \phantom{-}0.160(20)  \\
0.8 & \phantom{-}2.66(12)  & -1.31(12)  & \phantom{-}1.846(71) & \phantom{-}1.430(62)  & -0.326(12)  & -0.352(13)  & \phantom{-}1.97(14)  & -0.826(27) & \phantom{-}1.391(52) & \phantom{-}0.228(17)  \\
0.9 & \phantom{-}2.31(23)  & -1.20(13)  & \phantom{-}1.74(18)  & \phantom{-}1.463(55)  & -0.255(22)  & -0.284(14)  & \phantom{-}1.99(13)  & -0.808(42) & \phantom{-}1.427(69) & \phantom{-}0.292(18)  \\
1   & \phantom{-}3.106(35) & -1.288(71) & \phantom{-}1.85(13)  & \phantom{-}1.462(87)  & -0.180(41)  & -0.202(53)  & \phantom{-}1.65(24)  & -0.720(40) & \phantom{-}1.43(13)  & \phantom{-}0.341(17)  \\
\bottomrule\bottomrule
\end{tabular}
\caption{
    Values of the energy levels at $\beta = 6$. 
    These numbers correspond to the squares or points displayed in~\ref{fig:spectrum_beta6} and~\ref{fig:EcontPerMu_beta6} respectively.
}\label{tab:spectrum_beta6}
\end{table*}
 
\section{Complex Contour}\label{apx-sec:complex_contour}
Here we provide a short explanation for our choice of imaginary offset. 

In lattice field theory it has been known for a while that a contour deformation to the tangent plane of the main saddle point of the action, i.e.~the one with the greatest statistical weight, reduces the sign problem.
This point in $\mathbb{C}^\abs{\Lambda}$ fulfils $\left( \partial_{x,t}S[\phi]\right) \vline_{\phi=\phi_{c}} =0$. For the Hubbard model this tangent plane turns out to be parallel to the real axis due to symmetry, hence we are talking about an imaginary shift. Intuitively this improvement makes sense, because the integration manifold would be closer to the Lefschetz Thimbles.
The novelty of our recently developed contour deformation is the expansion of the action around said saddle point making it an effective action. This follows a standard practice in QFT and is equivalent to taking into account one-particle irreducible diagrams. We call this the next to leading order approximation (NLO). 
Because the linear term vanishes we expand until second order and get
\begin{equation}\label{eq:NLO}
	S_{\rm eff}[\phi_{c}]=S[\phi_{c}]+\frac{1}{2}\log\det\mathbb H_{S[\phi_{c}]}\ .
\end{equation}
as the new function to be minimized, where $\mathbb H$ is the hessian. This can be done numerically along the imaginary axis, i.e. $\phi_{{c}}=i\phi_1$.
By including the expansion we take into account the curvature of a saddle point, which shifts the classical (tangent) offset towards the optimal sign minimizing plane. We observe only small ranges of $\mu$ where it performs worse due to over-correction of steep regions in the action landscape.

Further details on the derivation and other optimizations can be found in~\cite{gantgen2024fermionic,rodekamp2024theory,gantgen2024reducing}.
     
\end{document}